\documentclass[12pt, preprint]{aastex}

\usepackage{natbib}
\usepackage{graphicx}
\usepackage{amsmath}
\usepackage{amssymb}
\usepackage{mathrsfs}
\usepackage[utf8]{inputenc}
\usepackage{longtable}

\shorttitle{Departure Coefficients for CRRLs }
\shortauthors{Salgado et al.}

\begin{document}
\title{Low Frequency Carbon Radio Recombination Lines I: Calculations of Departure Coefficients}

\author{F.~Salgado\altaffilmark{1}, L.~K.~Morabito\altaffilmark{1}, J.~B.~R.~Oonk\altaffilmark{1,2}, P.~Salas\altaffilmark{1}, M.~C.~Toribio\altaffilmark{1},
H.~J.~A.~R\"ottgering\altaffilmark{1}, A.~G.~G.~M., Tielens\altaffilmark{1}}

\altaffiltext{1}{Leiden Observatory, University of Leiden, P. O. Box 9513, 2300 RA Leiden, Netherlands}
\altaffiltext{2}{Netherlands Institute for Radio Astronomy (ASTRON), Postbus 2, 7990 AA Dwingeloo, The Netherlands}

\begin{abstract}
In the first paper of this series, we study the level population problem of recombining carbon ions. We focus our study on high quantum numbers
anticipating observations of Carbon Radio Recombination Lines to be carried out by the LOw Frequency ARray (LOFAR).
We solve the level population equation including angular momentum levels with updated collision rates up to high principal quantum numbers.
We derive departure coefficients by solving the level population equation in the hydrogenic approximation and including low temperature dielectronic
recombination effects. Our results in the hydrogenic approximation agree well with those of previous works. When comparing our results
including dielectronic recombination we find differences which we ascribe to updates in the atomic physics (e.g., collision rates) and to
the approximate solution method of the statistical equilibrium equations adopted in previous studies. A comparison with observations is discussed
in an accompanying article, as radiative transfer effects need to be considered.
\end{abstract}

\maketitle

\section{Introduction}

The interplay of stars and their surrounding gas leads to the presence of distinct phases in the interstellar medium (ISM) of galaxies (e.g. \citealt{field1969,mckee1977}).
Diffuse atomic clouds (the Cold Neutral Medium, CNM) have densities of about $50~\mathrm{cm^{-3}}$ and temperatures of about $80~\mathrm{K}$, where atomic hydrogen
is largely neutral but carbon is singly ionized by photons with energies between $11.2~\mathrm{eV}$ and $13.6~\mathrm{eV}$. The warmer ($\sim8000~\mathrm{K}$) and
more tenuous ($\sim0.5~\mathrm{cm^{-3}}$) intercloud phase is heated and ionized by FUV and EUV photons escaping from HII regions \citep{wolfire2003},
usually referred to as the Warm Neutral medium (WNM) and Warm Ionized Medium (WIM). The phases of the ISM are often globally considered to be in thermal equilibrium and in pressure balance \citep{savage1996, cox2005}. However,
the observed large turbulent width and presence of gas at thermally unstable, intermediate temperatures attests to the importance of heating by kinetic energy input.
In addition, the ISM also hosts molecular clouds, where hydrogen is in the form of $\mathrm{H_2}$ and self-gravity plays an important role. All of these
phases are directly tied to key questions on the origin and evolution of the ISM, including the energetics of the CNM, WNM and the WIM; the evolutionary
relationship of atomic and molecular gas; the relationship of these ISM phases with newly formed stars; and the conversion of their radiative and
kinetic power into thermal and turbulent energy of the ISM (e.g. \citealt{cox2005,elmegreen2004, scalo2004,mckee2007}).

The neutral phases of the ISM have been studied using optical and UV observations of atomic lines. These observations can provide the physical conditions but are
limited to pinpoint experiments towards bright background sources and are hampered by dust extinction \citep{snow2006}. At radio wavelengths,
dust extinction is not important and observations of the 21 cm hyperfine transition of neutral atomic hydrogen have been used to study the neutral phases
(e.g. \citealt{weaver1973,kalberla2005,heilesandtroland2003b}). On a global scale, these observations have revealed the prevalence of the two phase structure
in the interstellar medium of cold clouds embedded in a warm intercloud medium but they have also pointed out challenges to this theoretical view
\citep{kulkarni1987, kalberla2009}. It has been notoriously challenging to determine the physical characteristics (density, temperature) of the neutral structures
in the ISM as separating the cold and warm components is challenging (e.g. \citealt{heilesandtroland2003a}). In this context, Carbon radio recombination lines (CRRLs) provide a promising tracer of the neutral phases of the ISM (e.g. \citealt{peters2011, oonk2015a}). 

Carbon has a lower ionization potential (11.2~eV) than hydrogen (13.6~eV) and can be ionized by radiation fields in regions where hydrogen is largely neutral. Recombination of carbon ions with electrons to high Rydberg states will lead to CRRLs in the sub-millimeter to decameter wavelength range. Carbon radio recombination lines have been observed in the interstellar medium of our Galaxy towards two types of clouds: diffuse clouds (e.g.: \citealt{konovalenko1981, erickson1995, roshi2002, stepkin2007,oonk2014}) and photodissociation regions (PDRs), the boundaries of HII regions and their parent molecular clouds (e.g.: \citealt{natta1994, wyrowski1997, quireza2006}). The first low frequency (26.1 MHz) carbon radio recombination line was detected in absorption towards the supernova remnant Cas A by \citet{konovalenko1980} (wrongly attributed to a hyperfine structure line of ${^{14}}\mathrm{N}$, \citealt{konovalenko1981}). This line corresponds to a transition occurring at high quantum levels ($n=631$). Recently, \citet{stepkin2007} detected CRRLs in the range 25.5--26.5 MHz towards Cas A, corresponding to transitions involving levels as large as $n=1009$. 

Observations of low frequency carbon recombination lines can be used to probe the physical properties of the diffuse interstellar medium. However, detailed modeling
is required to interpret the observations. \citet{watson1980,walmsley1982} showed that, at low temperatures ($T_e \lesssim 100~\mathrm{K}$), electrons can recombine with carbon ions by simultaneously exciting the ${^2}P_{1/2}-{^2}P_{3/2}$~fine structure line, a process known as dielectronic recombination\footnote{ This process has been referred in the literature as dielectronic-like recombination or dielectronic capture by \citet{watson1980} to distinguish from the regular dielectronic recombination. Dielectronic capture refers to the capture of the electron in an excited $n$-state accompanied by simultaneous excitation of the ${^2}P_{1/2}$ core electron to the excited ${^2}P_{3/2}$ state. The captured electron can either auto ionize, collisional transferred to another state, or radiatively decay. Strictly speaking, dielectronic recombination refers to dieclectronic capture followed by stabilization. However, throughout this article we will use the term dielectronic recombination to refer to the same process as is common in the astronomical literature.}. Such recombination process occurs to high $n$ states, and can explain the behavior of the high $n$~CRRLs observed towards Cas A. \citet{walmsley1982} modified the code from \citet{brocklehurst1977} to include dielectronic recombination. \citet{payne1994} modified the code to consider transitions up to 10000 levels. All of these results assume a statistical distribution of the angular momentum levels, an assumption that is not valid at intermediate levels for low temperatures. Moreover, the lower the temperature, the higher the $n$-level for which that assumption is not valid.

The increased sensitivity, spatial resolution, and bandwidth of the Low Frequency ARray (LOFAR, \citealt{vhaarlem2013}) is opening the low frequency sky to
systematic studies of high quantum number radio recombination lines. The recent detection of high level carbon radio recombination lines using LOFAR
towards the line of sight of Cas A \citep{asgekar2013}, Cyg A \citep{oonk2014}, and the first extragalactic detection in the starburst galaxy M82 \citep{morabito2014b}
illustrate the potential of LOFAR for such studies. Moreover, pilot studies have demonstrated that surveys of low frequency radio recombination lines of
the galactic plane are within reach, providing a new and powerful probe of the diffuse interstellar medium. These new observations have motivated us to reassess some
of the approximations made by previous works and to expand the range of applicability of recombination line theory in terms of physical parameters. In addition,
increased computer power allows us to solve the level population problem considering a much larger number of levels than ever before. Furthermore,
updated collisional rates are now available \citep{vrinceanu2012}, allowing us to explicitly consider the level population of quantum angular momentum sub-levels
to high principal quantum number levels.
Finally, it can be expected that the Square Kilometer Array, SKA, will further revolutionize our understanding of the low frequency universe
with even higher sensitivity and angular resolution \citep{oonk2015a}.

In this work, we present the method to calculate the level population of recombining ions and provide some exemplary results applicable to low temperature diffuse
clouds in the ISM. In an accompanying article (\citealt{salgado2016}, from here on Paper II), we will present results specifically geared
towards radio recombination line studies of the diffuse interstellar medium. In Section~\ref{section_method}, we introduce the problem of level
populations of atoms and the methods to solve this problem for hydrogen and hydrogenic carbon atoms. We also present the rates used in this work
to solve the level population problem. In Section~\ref{section_results}, we discuss our results focusing on hydrogen and carbon atoms. We compare
our results in terms of the departure coefficients with previous results from the literature. In Section~\ref{section_conclusions}, we summarize
our results and provide the conclusions of the present work.

\section{Theory}\label{section_method}
A large fraction of our understanding of the physical processes in the Universe comes from observations of atomic lines in astrophysical plasmas.
In order to interpret the observations, accurate models for the level population of atoms are needed as the strength (or depth) of an emission (absorption)
line depends on the level populations of atoms. Here, we summarize the basic ingredients needed to build level population models and provide a basic description of
the level population problem. We begin our discussion by describing the line emission and absorption coefficients ($j_\nu$ and $k_\nu$, respectively),
which are given by \citep{shaver1975, gordon2009}:
\begin{eqnarray}
j_\nu &=& \frac{h \nu}{4 \pi} A_{n'n} N_{n'} \phi(\nu),\\
k_\nu &=& \frac{h \nu}{4 \pi} \left( N_{n} B_{nn'}-N_{n'} B_{n'n}  \right) \phi(\nu),
\end{eqnarray}
\noindent where $h$ is the Planck constant, $N_{n'}$ is the level population of a given upper level ($n'$) and $N_{n}$ is the level
population of the lower level ($n$); $\phi(\nu)$ is the line profile, $\nu$ is the frequency of the transition and $A_{n'n}$,
$B_{n'n}(B_{nn'})$ are the Einstein coefficients for spontaneous and stimulated emission (absorption)
\footnote{We provide the formulation to obtain the values for the rates in Appendix C.}, respectively.

Under local thermodynamic equilibrium (LTE) conditions, level populations are given by the Saha-Boltzmann equation (e.g. \citealt{brocklehurst1971}):
\begin{eqnarray}
N_{nl}(LTE)&=&N_e N_{ion}\left(\frac{h^2}{2 \pi m_e k T_e}\right)^{3/2} \frac{\omega_{nl}}{2\omega_i} e^{\chi_n}, \chi_n=\frac{hc Ry Z^2}{n^2kT_e},
\end{eqnarray}
\noindent where $T_e$ is the electron temperature, $N_e$~is the electron density in the nebula, $N_{ion}$~is the ion density, $m_e$~is the electron mass, $k$ is the Boltzmann constant,
$h$ is the Planck constant, $c$ is the speed of light and $Ry$ is the Rydberg constant; $\omega_{nl}$~is the statistical weight of the level $n$ and
angular quantum momentum level $l$~[$\omega_{nl}=2(2l+1)$, for hydrogen], and $\omega_i$ is the statistical weight of the parent ion. The factor
$\left({h^2}/{2 \pi m_e k T_e}\right)^{1/2}$~is the thermal de~Broglie wavelength, $\Lambda(T_e)$, of the free electron
\footnote{$\Lambda(T_e)^3\approx4.14133\times10^{-16}~T_e^{-1.5}~\mathrm{cm^3}$.}. In the most general case,
lines are formed under non-LTE conditions and the level population equation must be solved in order to properly model the line properties as a function
of quantum level ($n$).

Following e.g. \citet{seaton1959a} and \citet{brocklehurst1970}, we present the results of our modeling in terms of the departure coefficients ($b_{nl}$), defined by:
\begin{eqnarray}\label{eqn_bn}
b_{nl}=\frac{N_{nl}}{N_{nl}(LTE)},
\end{eqnarray}
\noindent and $b_n$~values are computed by taking the weighted sum of the $b_{nl}$ values:
\begin{eqnarray}\label{eqn_bn2}
b_n=\sum_{l=0}^{n-1} \left(\frac{2l+1}{n^2}\right)b_{nl},
\end{eqnarray}
\noindent note that, at a given $n$, the $b_{nl}$~values for large $l$~levels influence the final $b_n$~value the most due to the statistical
weight factor. At low frequencies stimulated emission is important \citep{goldberg1966} and we introduce the correction factor for stimulated emission
as defined by \citet{brocklehurst1972}:
\begin{eqnarray}\label{eq_beta}
\beta_{n,n'}=\frac{1-\left(b_{n'}/b_n\right) \exp(-h \nu/k T_e)}{1-\exp(-h \nu/k T_e)},
\end{eqnarray}
\noindent unless otherwise stated the $\beta_n$~presented here correspond to $\alpha$~transitions ($n'=n+1\rightarrow n$).
The description of the level population in terms of departure coefficients is convenient as it reduces the level population problem to a more
easily handled problem as we will show in Section \ref{section_levelpop}.

\subsection{Level Population of Carbon Atoms under Non-LTE Conditions}\label{section_levelpop}
The observations of high $n$ carbon recombination lines in the ISM motivated \citet{watson1980} to study the effect on the level population of dielectronic recombination
and its inverse process (autoionization) in low temperature ($T_e\lesssim 100~\mathrm{K}$) gas. \citet{watson1980} used 
$l$-changing collision rates \footnote{We use the term $l$-changing collision rates to refer to collisions rate that induce a transition from state $nl$ to $nl\pm1$} from \citet{jacobs1978} and concluded that for levels $n\approx250-300$, dielectronic recombination of carbon ions
can be of importance. In a later work, \citet{walmsley1982} used collision rates from \citet{dickinson1981} and estimated a value for which
autoionization becomes more important than angular momentum changing rates. The change in collision rates led them to conclude
that the influence of dielectronic recombination on the $b_n$~values is important at levels $n\gtrsim300$. Clearly, the results are sensitive
to the choice of the angular momentum changing rates. Here, we will explicitly consider $l$-sublevels when solving the level population equation.

The dielectronic recombination and autoionization processes affect only the C$^+$ ions in the ${^2}P_{3/2}$ state, therefore we treat the level population for the two
ion cores in the ${^2}P_{1/2}$ states separately in the evaluation of the level population \citep{walmsley1982}.
The equations for carbon atoms recombining to the ${^2}P_{3/2}$~ion core population have to include terms describing dielectronic recombination ($\alpha_{nl}^d$) and autoionization ($A_{nl}^a$), viz.:
\begin{eqnarray}\label{eqn_llevelpopc}
b_{nl}\left[\sum\limits_{n' < n}\sum_{l'=l\pm1}A_{nln'l'}+\sum\limits_{n' \neq n}{(B_{nln'l'}I_\nu + C_{nln'l'})}+\sum_{l'=l\pm1}C_{nlnl'}+ A_{nl}^a +C_{nl,i}\right] &=&\nonumber \\
 \sum\limits_{n' > n}\sum_{l'=l\pm1}{b_{n'l'} \frac{\omega_{n'l'}}{\omega_{nl}} e^{\Delta \chi_{n'n}} A_{n'l'nl}}+\sum\limits_{n' \neq n}\sum_{l'=l\pm1}{b_{n'l'} \frac{\omega_{n'l'}}{\omega_{nl}} e^{\Delta \chi_{n'n}}(B_{n'l'nl}I_\nu +C_{n'l'nl})}+\nonumber\\ 
+\sum_{l'=l\pm1} b_{nl'} \left(\frac{\omega_{nl'}}{\omega_{nl}}  \right)C_{nl'nl}+\frac{N_e N_{3/2}^+}{N_{nl}(LTE)}(\alpha_{nl}+ C_{i,nl})+ \frac{N_e N_{1/2}^+}{N_{nl}(LTE)}\alpha_{nl}^d.
\end{eqnarray}
\noindent The left hand side of Equation~\ref{eqn_llevelpopc} describes all the processes that take an electron out of the $nl$-level, and the right hand side
the processes that add an electron to the $nl$ level; $A_{nln'l'}$ is the coefficient for spontaneous emission, $B_{nln'l'}$ is the coefficient for stimulated emission
or absorption induced by a radiation field $I_\nu$; $C_{nln'l'}$ is the coefficient for energy changing collisions (i.e. transitions with $n\neq n'$), $C_{nlnl'}$ is the coefficient 
for $l$-changing collisions; $C_{nl,i}$ ($C_{i,nl}$) is the coefficient for collisional ionization (3-body recombination) and $\alpha_{nl}$ is the coefficient for
radiative recombination. A description of the coefficients entering in Equation~\ref{eqn_llevelpopc} is given in Section~\ref{section_rates} and in further detail
in the Appendix. The level population equation is solved by finding the values for the departure coefficients. The level population for carbon ions recombining to the
${^2}P_{1/2}$~level is hydrogenic and we solve for the departure coefficients ($b_{nl}^{1/2}$) using Equation~\ref{eqn_llevelpopc}, but ignoring the coefficients
for dielectronic recombination and autoionization.

After computing the $b_{nl}^{1/2}$ and $b_{nl}^{3/2}$, we compute the departure coefficients ($b_n^{1/2}$ and $b_n^{3/2}$) for both parent ion populations by summing
over all $l$-states (Equation~\ref{eqn_bn2}). The final departure coefficients for carbon are obtained by computing the weighted average of both ion cores:
\begin{eqnarray}\label{eq_bnfinal}
b_n^{final} &=& \frac{b_n^{1/2}+b_n^{3/2} \left[N^+_{3/2}/N^+_{1/2}\right]}{1+\left[N^+_{3/2}/N^+_{1/2}\right]}.
\end{eqnarray}

\begin{figure}[!ht]
\includegraphics[width=0.5\columnwidth]{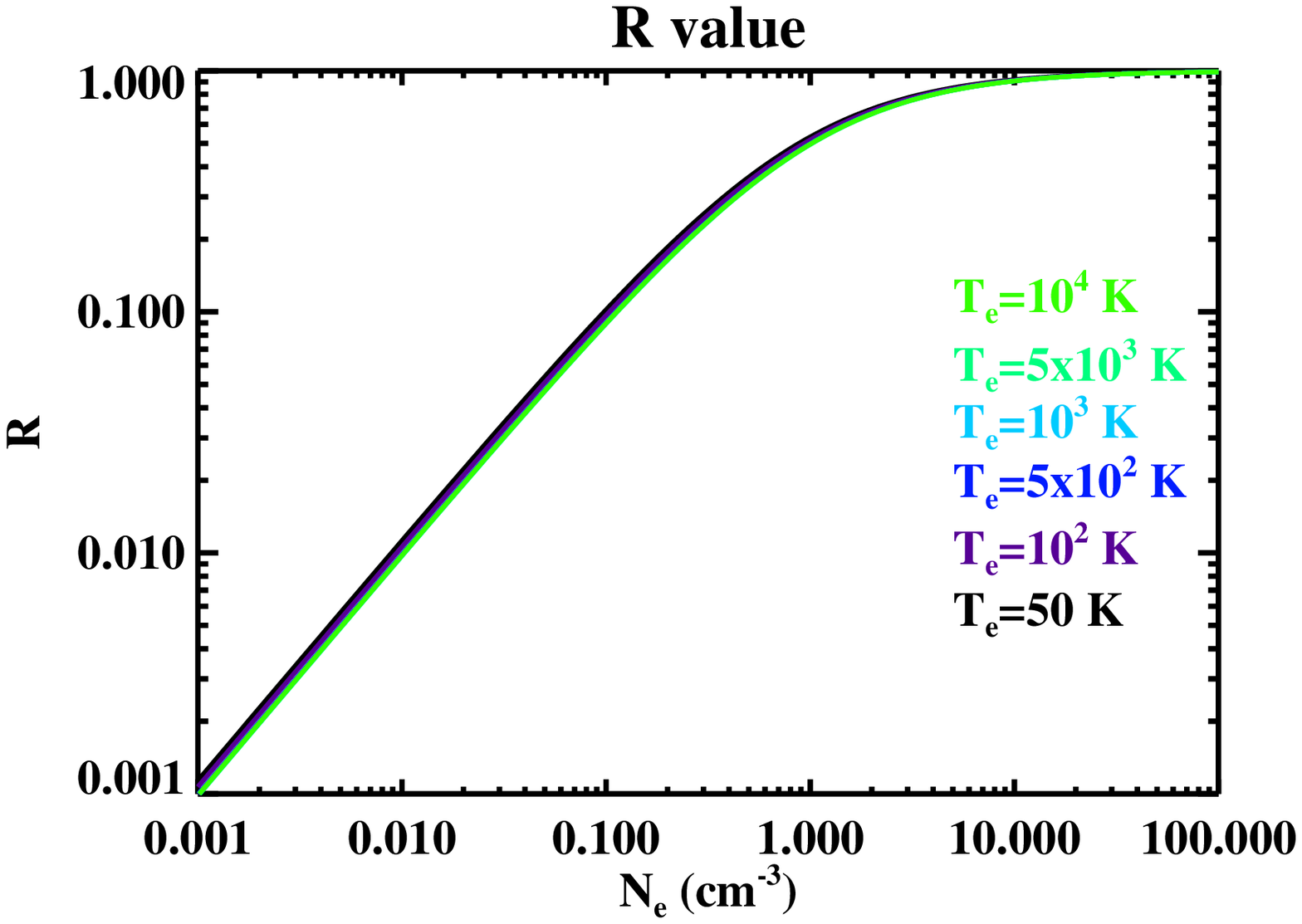}
\includegraphics[width=0.5\columnwidth]{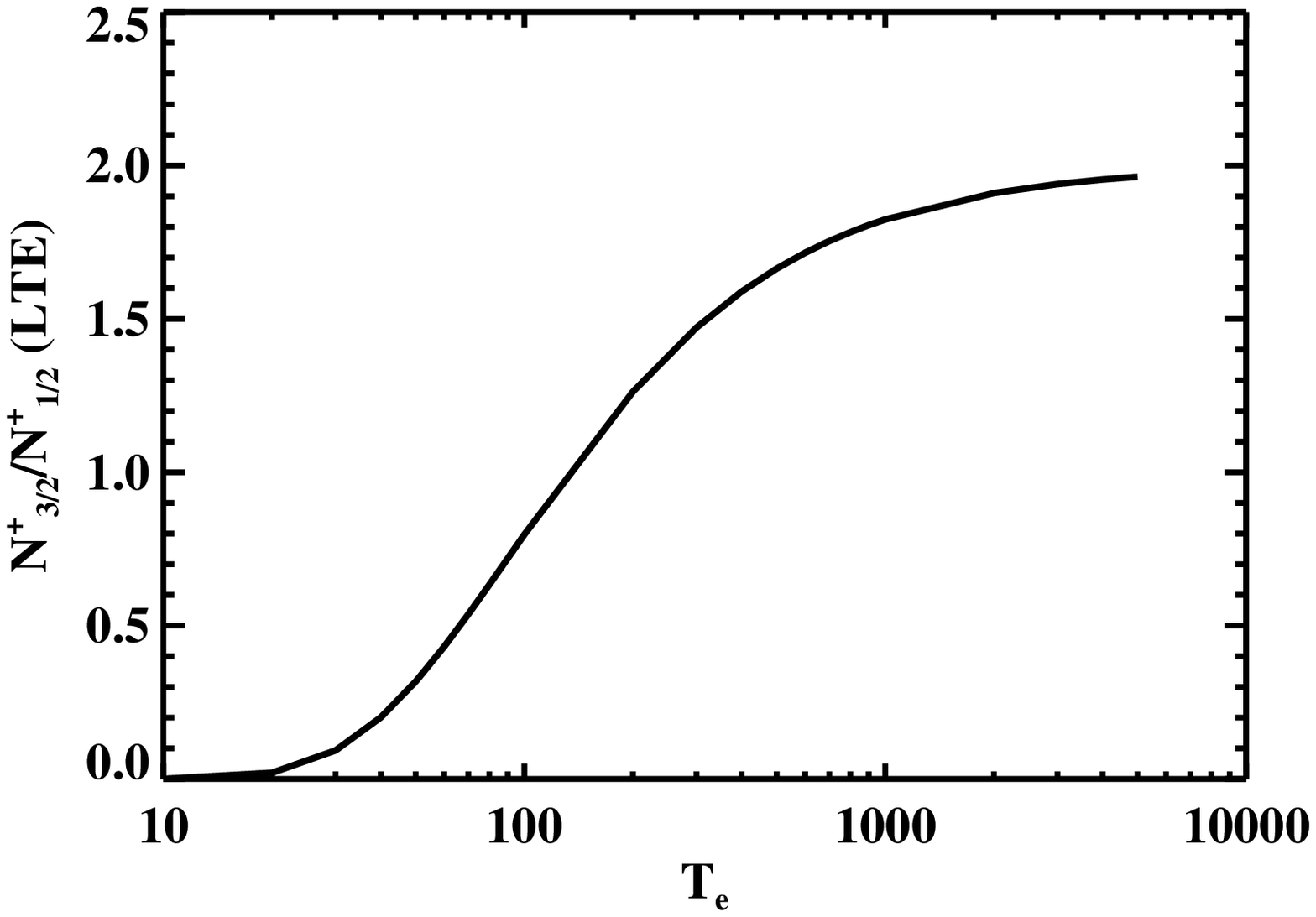}
\caption{Left panel: $R$ value as a function of electron temperature, in a range of densities. The $R$ value is nearly independent of temperature,
and for $N_e > 10~\mathrm{cm^{-3}}$, $R\approx1$. Right panel: ion ``LTE'' ratios as a function of $T_e$, independent of density.\label{fig_rvalue}}
\end{figure}

Note that, in order to obtain the final departure coefficients, the relative population of the parent ion cores is needed. Here, we assume that the population ratio of the two ion cores
$N^+_{3/2}$~to $N^+_{1/2}$~is determined by collisions with electrons and hydrogen atoms. This ratio can be obtained using \citep{ponomarev1992,payne1994}:
\begin{eqnarray}\label{eqpayner}
R &=& \frac{N_{3/2}^{+}/N_{1/2}^{+}}{N_{3/2}^{+}(LTE)/N_{1/2}^{+}(LTE)} \\
  &=& \frac{N_e\gamma_e+N_H\gamma_H}{N_e\gamma_e+N_H\gamma_H+A_{3/2,1/2}},
\end{eqnarray}
\noindent where $\gamma_e = 4.51\times 10^{-6}~T_e^{-1/2}~\mathrm{cm^{-3}~s^{-1}}$~is the de-excitation rate due to collisions with electrons,
$\gamma_H = 5.8 \times 10^{-10}~T_e^{0.02}~\mathrm{cm^{-3}~s^{-1}}$~is the de-excitation rate due to collisions with hydrogen atoms \citep{payne1994}
\footnote{\citet{payne1994} used rates from \citet{tielens1985}, based on \citet{launay1977} for collisions with hydrogen atoms and \citet{hayes1984}
for collisions with electrons. Newer rates are available for collisions with electrons \citep{wilson2002} and hydrogen atoms \citep{barinovs2005},
but the difference in values is negligible.}, $N_\mathrm{H}$ is the atomic hydrogen density and $A_{3/2,1/2}=2.4 \times 10^{-6}~\mathrm{s^{-1}}$~is
the spontaneous radiative decay rate of the core. In this work, we have ignored collisions with molecular hydrogen, which should be included for high
density PDRs. Collisional rates for H$_2$ excitation of C$^+$ have been calculated by \citet{flower1988}. In the cases of interest here, the value
of $R$~is dominated by collisions with atomic hydrogen. We recognize that the definition of $R$ given in Equation~\ref{eqpayner} is related
to the critical density ($N_{cr}$) of a two level system by $R=1/(1+ N_{cr}/N_X)$ where $N_X$ is the density of the collisional partner (electron or hydrogen).
The LTE ratio of the ion core is given by the statistical weights of the levels and the temperature ($T_e$) of the gas:
\begin{eqnarray}\label{eqn32n12lte}
\frac{N_{3/2}^+(LTE)} {N_{1/2}^+(LTE)} &=& \frac{g_{3/2}}{g_{1/2}} e^{-\Delta E/k T_e},
\end{eqnarray}
\noindent where $g_{3/2}=4,~g_{1/2}=2$ are the statistical weights of the fine structure levels and $\Delta E=92~\mathrm{K}$~is the energy difference of the fine structure transition.
The LTE level population ratio as a function of temperature is shown in Figure~\ref{fig_rvalue}, illustrating the strong dependence on temperature of this value.
At densities below the critical density ($\approx300~\mathrm{cm^{-3}}$ for collisions with H), the fine structure levels fall out of LTE and the value for $R$ becomes very small (Figure~\ref{fig_rvalue}). Note that $R$ is not very
sensitive to the temperature.

With the definition of $R$ given above, the final departure coefficient can be written as \citep{ponomarev1992}:
\begin{eqnarray}\label{equation_bnfinal2}
b_n^{final} &=& \frac{b_n^{1/2}+b_n^{3/2} R\left[N^+_{3/2}/N^+_{1/2}\right]_{LTE}}{1+R \left[N^+_{3/2}/N^+_{1/2}\right]_{LTE}}.
\end{eqnarray}
\noindent The final departure coefficient is the value that we are interested in to describe CRRLs.

\subsection{Numerical Method}\label{section_nummethod}
Having described how to derive the $b_{n}^{final}$, now we focus on the problem of obtaining the departure coefficients for both ion cores from the level population equation.
We use the same procedure to obtain the departure coefficients for both parent ion cores, as the only difference in the level population equation for the ${^2}P_{3/2}$
and the ${^2}P_{1/2}$ cores is the inclusion of dielectronic recombination and autoionization processes. We will refer as $b_{nl}$ and $b_n$ without making a distinction
between the ${^2}P_{3/2}$ and ${^2}P_{1/2}$ in this subsection.

We follow the methods described in \citet{brocklehurst1971} and improved in \citet{hummer1987} to solve the level population equation in an iterative manner.
First, we solve the level population equation by assuming that the $l$ sublevels are in statistical equilibrium, i.e. $b_n=b_{nl}$ for all $l$~sublevels. We refer
to this approach as the $n$-method (see Appendix B). Second, we used the previously computed values to determine the coefficients on the right hand side of
Equation~\ref{eqn_llevelpopc} that contain terms with $n'\neq n$. Thus, the level population equation for a given $n$ is a tridiagonal equation on the $l$ sublevels
involving terms of the type $l\pm1$. This tridiagonal equation is solved for the $b_{nl}$ values (further details are given in Appendix B). The second step of
this procedure is repeated until the difference between the computed departure coefficients is less than 1\%.

We consider a fixed maximum number of levels, $n_{max}$, equal to 9900. We make no explicit assumptions on the asymptotic behavior of the $b_n$~for larger
values of $n$. Therefore, no fitting or extrapolation is required for large $n$. The adopted value for $n_{max}$ is large enough for the asymptotic limit
-- $b_n \rightarrow 1$ for $n>n_{max}$ -- to hold even at the lowest densities considered here. For the $nl$-method, we need to consider all $l$ sublevels up to a
high level ($n\sim1000$). For levels higher than this critical $n$ level ($n_{crit}$), we assume that the $l$ sublevels are in statistical equilibrium. In our
calculations $n_{crit}=1500$, regardless of the density.

\subsection{Rates Used in this Work}\label{section_rates}
In this section, we provide a brief description of the rates used in solving the level populations. Further details and the mathematical formulations
for each rate are given in Appendices C, D, E and F. Accurate values for the rates are critical to obtain meaningful departure coefficients when solving the level
population equation (Equation~\ref{eqn_llevelpopc}). Radiative rates are known to high accuracy ($< 1\%$) as they can be computed from first principles. On the
other hand, collision rates at low temperatures are more uncertain ($\sim 20\%$, \citealt{vriens1980}).

\subsubsection{Einstein A and B coefficient}
The Einstein coefficients for spontaneous and stimulated transitions can be derived from first principles.
We used the recursion formula described in \citet{storey1991} to obtain the values for the Einstein $A_{nln'l'}$~coefficients.
To solve the $n$~method (our first step in solving the level population equation) we require the values for $A_{nn'}$, which can be easily obtained by summing the $A_{nln'l'}$:
\begin{eqnarray}\label{avgacoeff}
A_{n n'}&=&\frac{1}{n^2}\sum\limits_{l'=0}\limits^{n-1} \sum\limits_{l=l'\pm1}(2 l+ 1) A{_{nl}}{_{n'l'}}.
\end{eqnarray}
The mathematical formulation to obtain values for spontaneous transitions is detailed in Appendix \ref{app_einsa}.

The coefficients for stimulated emission and absorption ($B_{nn'}$) are related to the $A_{nn'}$~coefficients by:
\begin{eqnarray}
B_{nn'}&=& \frac{c^2}{2 h \nu^3} A_{nn'},\\
B_{n'n}&=& \left(\frac{n}{n'}\right)^2 B_{nn'}.
\end{eqnarray}

\subsubsection{Energy changing collision rates}
\begin{figure}[!ht]
\includegraphics[width=0.5\columnwidth]{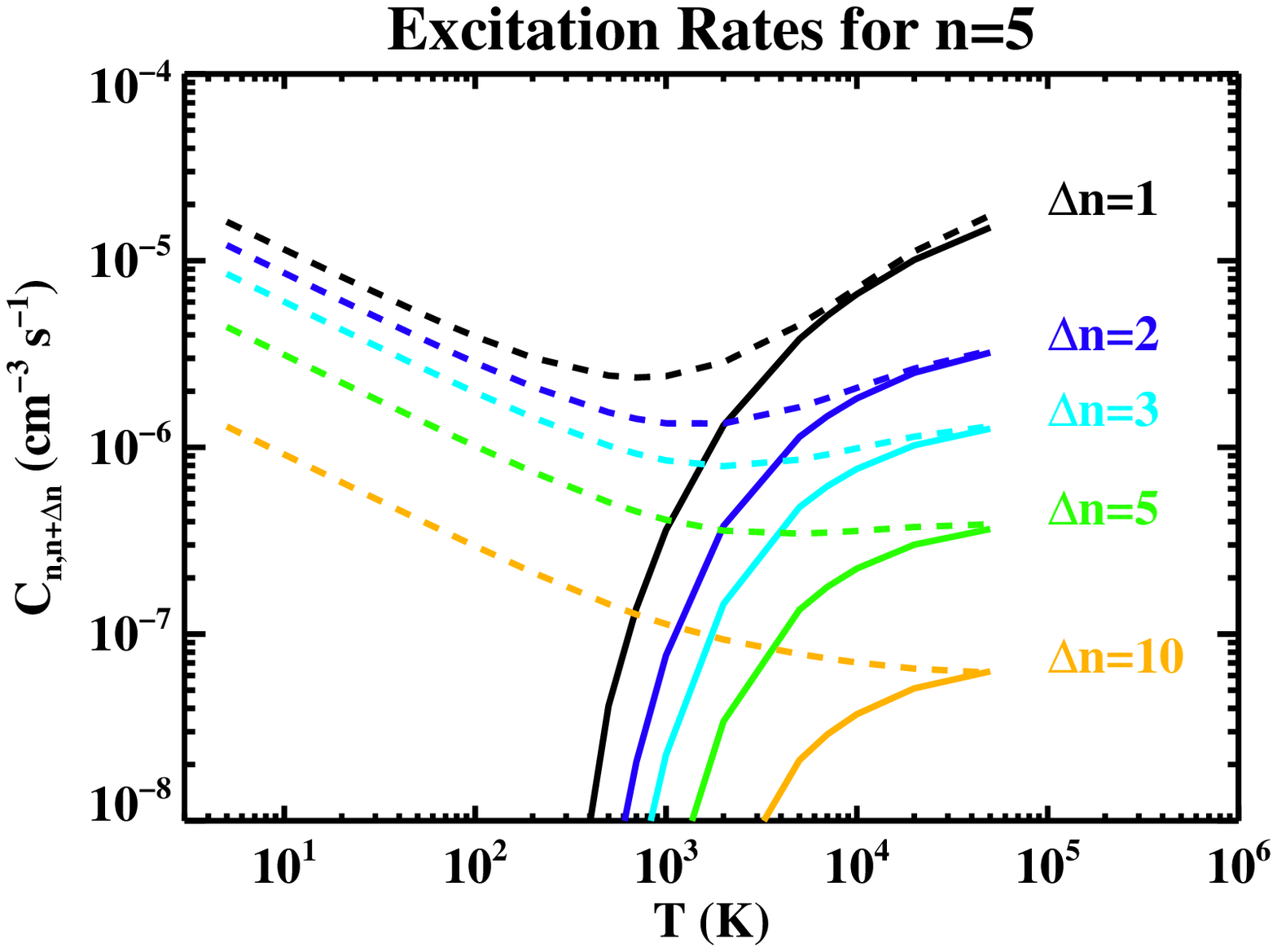}
\includegraphics[width=0.5\columnwidth]{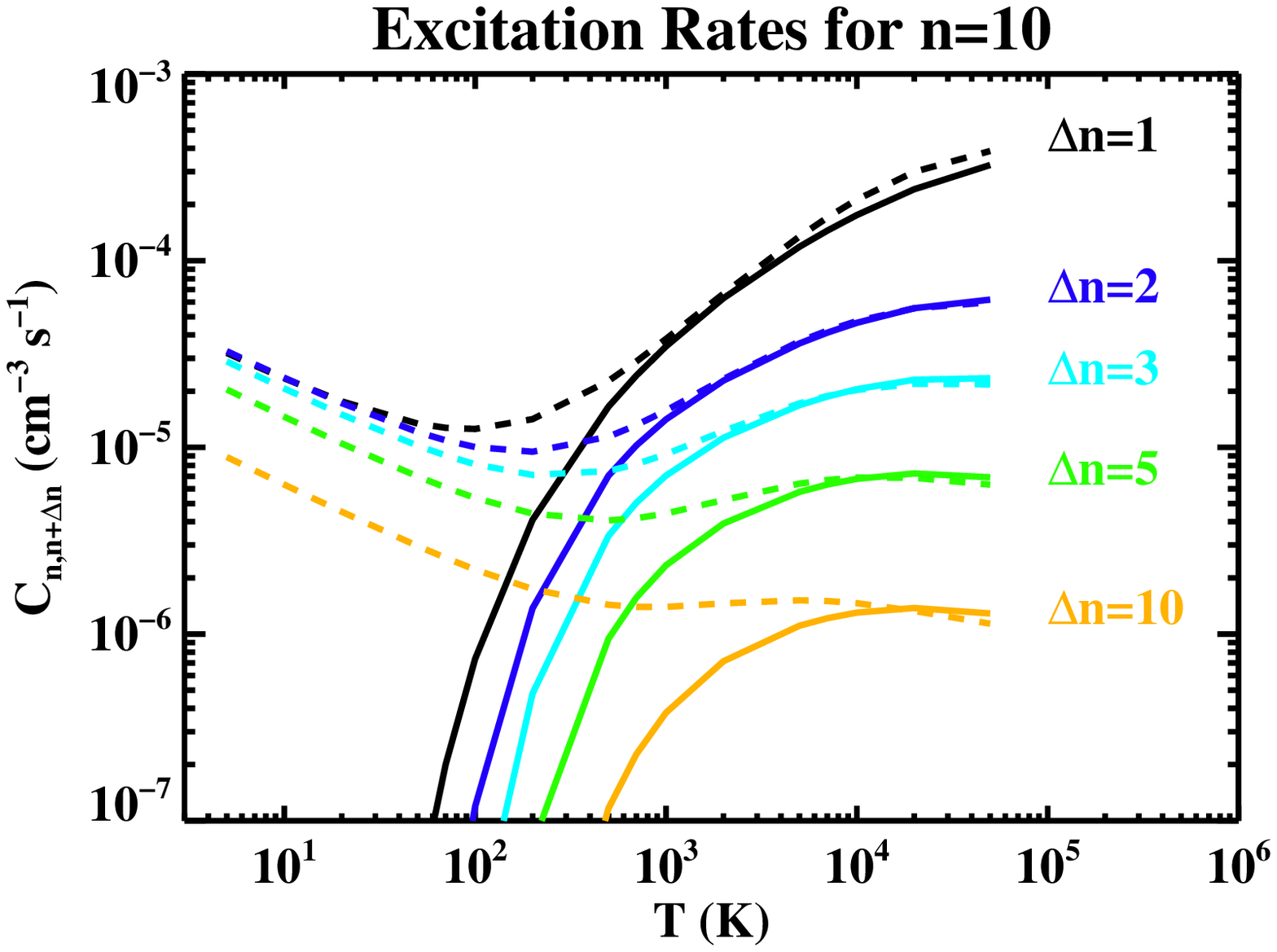}
\includegraphics[width=0.5\columnwidth]{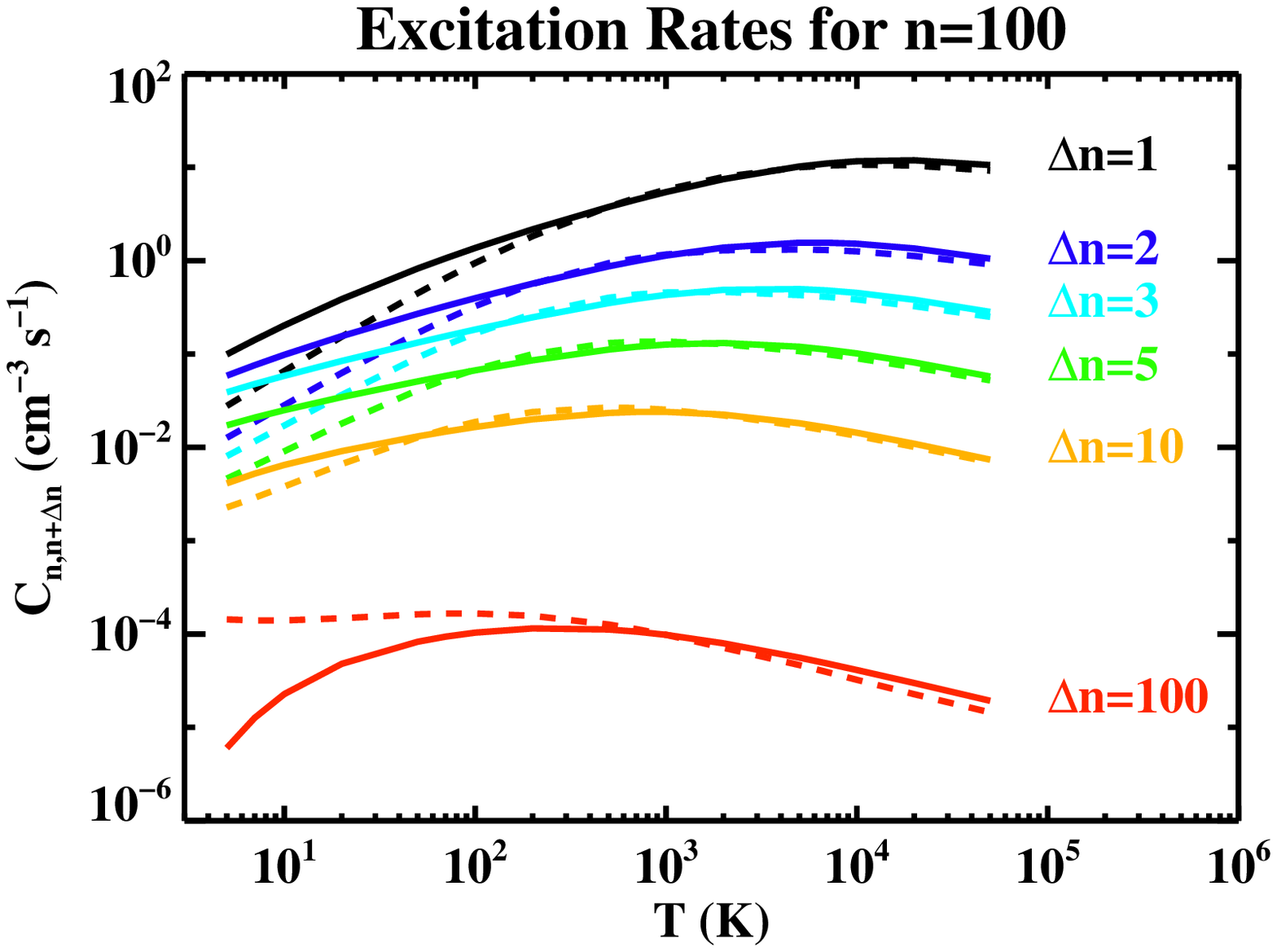}
\includegraphics[width=0.5\columnwidth]{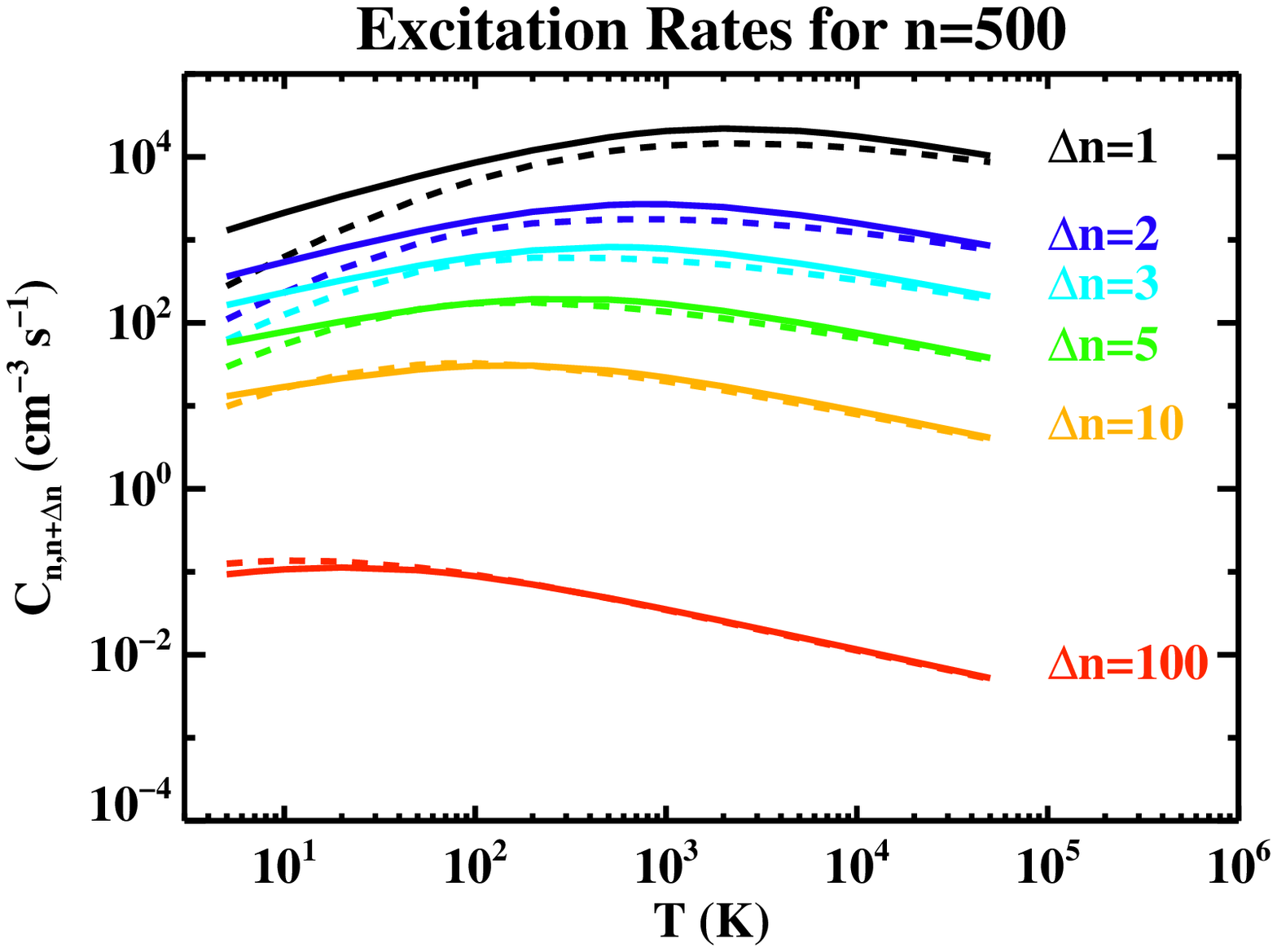}
\caption{Comparison of energy changing collision rates. The dashed lines correspond to the \citet{gee1976} rates while the solid lines are from \citet{vriens1980}.
Large differences between \citet{gee1976} and \citet{vriens1980} can be seen at low $T_e$~and at low $n$~levels.
As is well known, transitions with $\Delta n=1$~dominate. The difference between $\Delta n>1$~and $\Delta n=1$~rates
is less at lower $T_e$.\label{fig_col}}
\end{figure}

In general, energy changing collisions are dominated by the interactions of electrons with the atom. The interaction of an electron with an atom can induce
transitions of the type:
\begin{eqnarray}
X_{nl} + e^-\rightleftarrows X_{n'l'} + e^-,
\end{eqnarray}
\noindent with $n'\neq n$ changing the distribution of electrons in an atom population. \citet{hummer1987} used the formulation of \citet{percival1978}.
The collision rates derived by \citet{percival1978} are essentially the same as \citet{gee1976}.
However, the collision rates from \citet{gee1976} are not valid for the low temperatures of interest here. Instead, we use collision rates from \citet{vriens1980}.
We note that at high $T_e$~and for high $n$~levels, the Bethe (Born) approximation holds and values of the rates from \citet{vriens1980} differ by less than 20\%
when compared to those from \citet{gee1976}. The good agreement between the two rates is expected since the results from \citet{vriens1980} are based on \citet{gee1976}.
On the other hand, at low $T_e$~and for low $n$~levels values the two rates differ by several orders of magnitude and, indeed, the \citet{gee1976} values are too high to
be physically realistic. A comparison of the rates for different values of $T_e$~and $n\rightarrow n+ \Delta n$ transitions is shown in Figure~\ref{fig_col}.
We explore the effects of using \citet{vriens1980} rates on the $b_n$ values in Section~\ref{section_comparison}.

The inverse rates are obtained from detailed balance:
\begin{eqnarray}
C_{n'n}= \left(\frac{n}{n'}\right)^2 e^{\chi_n-\chi_{n'}} C_{nn'}.
\end{eqnarray}

In order to solve the $nl$-method, rates of the type $C_{nln'l'}$ with $n\neq n'$ are needed. Here, the approach of \citet{hummer1987} is followed and
the collision rates are normalized by the oscillator strength of the transitions (Equation~5 in \citealt{hummer1987}). Only transitions
with $\Delta l=1$~were included as these dominate the collision process \citep{hummer1987},.

\subsubsection{Angular momentum changing collision rates}
For low $n$~levels, the $l$~level population has to be explicitly calculated. Moreover, for the dielectronic recombination process, the angular
momentum changing collisions set the value for which the dielectronic recombination process is important, and transitions of the type:
\begin{eqnarray}
X_{nl} + \mathrm{C^+}\rightleftarrows X_{nl\pm1} + \mathrm{C^+}
\end{eqnarray}
\noindent must be considered.  In general, collisions with ions are more important than collisions with electrons. Here, for simplicity, we adopt
that $\mathrm{C}^+$ is the dominant cation.

\citet{hummer1987} used $l$-changing collision rates from \citet{pengellyandseaton1964} which are computed iteratively for a given $n$ level starting at $l=0$ or $l=n-1$. However,
as pointed out by \citet{hummer1987} and \citet{brocklehurst1971}, the values for the $l$-changing rates obtained by starting the iterations at $l=0$ differ from those obtained when starting at $l=n-1$.
Moreover, averaging the $l$-changing rates obtained by the two different initial conditions leads to an oscillatory behavior of the rates that depends on $l$ \citep{brocklehurst1970}.
\citet{hummer1987} circumvented this problem by normalizing the value of the rates by the oscillator strength (Equation~4 in \citealt{hummer1987}). 
In addition, at high $n$~levels and high densities the values for $C_{nln'l'}$ can become negative (Equation~43 in \citealt{pengellyandseaton1964}).
This poses a problem when studying the level population of carbon atoms at the high $n$~levels of interest in the present work\footnote{We note that
this was not a problem for \citet{hummer1987}, since they assumed an statistical distribution of the $l$~levels for high $n$.}.
The more recent study of \citet{vrinceanu2012} provides a general formulation to obtain the value of $l$-changing transition rates.
These new rates use a much smaller cut-off radius of the probability of the transition for large impact parameters. Furthermore, the rates from \citet{vrinceanu2012}
are well behaved over a large range of temperature and densities and they do not exhibit the oscillatory behavior with $l$ sublevel shown by the
\citet{pengellyandseaton1964} rates. Therefore, we use the \citet{vrinceanu2012} rates in this work. \citet{vrinceanu2012} derived the following expression,
valid for $n>10$~and $n \sqrt{T_e}<2.4\times 10^4~\mathrm{K^{1/2}}$:
\begin{eqnarray}
C_{nl\rightarrow nl+1} &=& 12 \sqrt{\pi} a_0^3 \left(2 \pi c Ry\right) \sqrt{\left( \frac{hc Ry}{k T_e}\right) \left( \frac{\mu}{m_e}\right)} n^4 \left[1-\left( \frac{l}{n} \right)^2 \left( \frac{2l+3}{2l+1} \right) \right].
\end{eqnarray}
\noindent where $a_0$ is the Bohr radius and $\mu$ is the reduced mass of the system.
Values for the inverse process are obtained by using detailed balance:
\begin{eqnarray}
C_{nl+1\rightarrow nl}&=& \frac{(2l+1)}{(2l+3)}C_{nl\rightarrow nl+1}.
\end{eqnarray}

We note that the $l$-changing collision rates obtained by using the formula from \citet{vrinceanu2012} can differ by a factor of six \citep{vrinceanu2012} with those using the \citet{pengellyandseaton1964} formulation. We discuss the effect on the final $b_n$~values in Section~\ref{section_comparison}, where we compare our results with those of \citet{storey1995} in the Hydrogenic approximation and with those of \citet{ponomarev1992} for carbon atoms.

\subsubsection{Radiative Recombination}
Radiative ionization occurs when an excited atom absorbs a photon with enough energy to ionize the excited electron.
The process can be represented as follows:
\begin{eqnarray}
X_{nl} + h\nu \rightleftarrows X^+ + e^-,
\end{eqnarray}
\noindent and the inverse process is radiative recombination.
We use the recursion relation described in \citet{storey1991} to obtain values for the ionization cross-section (Appendix \ref{app_radrec}). Values
for the radiative recombination ($\alpha_{nl}$) coefficients were obtained using the Milne relation and standard formulas (e.g. \citet{rybicki1986},
Appendix \ref{app_radrec}). The program provided by \citet{storey1991} only produces reliable values up to $n\sim500$ due to cancellation effects in
the iterative procedure. In order to avoid cancellation effects, the values computed here were obtained by working with logarithmic values in the
recursion formula. As expected, our values for the rates match those of \citet{storey1991} well.

For the $n$-method we require the sum of the individual $\alpha_{nl}$~values:
\begin{eqnarray}
\alpha_{n}=\sum_{l=0}^{n-1}\alpha_{nl}.
\end{eqnarray}
\noindent The averaged $\alpha_n$ values agree well with the approximated formulation of \citet{seaton1959a} to better than 5\%, validating our approach.

\subsubsection{Collisional ionization and 3-body recombination}
Collisional ionization occurs when an atom encounters an electron and, due to the interaction, a bound electron
from the atom is ionized. Schematically the process can be represented as:
\begin{eqnarray}
 X_n + e^- &\rightleftarrows& X^+ + e^- +e^-.
\end{eqnarray}
The inverse process is given by the 3-body recombination and the value for the 3-body recombination rate is obtained from detailed balance:
\begin{eqnarray}
C_{i,n}   &=& \frac{N_n(LTE)}{N_{ion} N_e} C_{n,i}, \nonumber \\
         &=& \left(\frac{h^2}{2\pi m_e k Te}\right)^{3/2} n^2 e^{\chi_n} C_{n,i},  \nonumber \\
         &=& \Lambda(T_e)^3 n^2 e^{\chi_n} C_{n,i}.
\end{eqnarray}
We used the formulation of \citet{brocklehurst1977} and compared the values with those from the formulation given by \citet{vriens1980}.
For levels above 100 and at $\mathrm{T_e} =10~\mathrm{K}$, the Brocklehurst \& Salem values are a factor of $\lesssim 2$ larger,
but the differences quickly decrease for higher temperatures. To obtain the $C_{nl,i}$ that are needed in the $nl$-method, we
followed \citet{hummer1987} and assumed that the rates are independent of the angular momentum. The mathematical formulation is
reproduced in the Appendix \ref{app_cni} for convenience of the reader.

\subsubsection{Dielectronic Recombination and Autoionization on Carbon Atoms}
The dielectronic recombination process involves an electron recombining into a level $n$~while simultaneously exciting
one of the bound electrons (left side of Equation~\ref{dielrec1}, below). This state ($X_n^*$) is known as an autoionizing state.
In this autoionizing state, the atom can stabilize either by releasing the recombined electron through autoionization (inverse process
of dielectronic recombination) or through radiative stabilization (right hand side of Equation~\ref{dielrec1}). Dielectronic recombination
and autoionization are only relevant for atoms with more than one electron.
\begin{eqnarray}\label{dielrec1}
X^+ + e^-\rightleftarrows X_{nl}^* \rightarrow X_{n'l'} + h\nu.
\end{eqnarray}
For C$^+$ recombination, at $T_e\sim100~\mathrm{K}$~free electrons in the plasma can recombine to a high $n$~level, and the kinetic energy is transfered
to the core of the ion, producing an excitation of the ${^2}P_{1/2}-{^2}P_{3/2}$ fine-structure level of the C$^+$ atom core
(which has a difference in energy $\Delta E=92~\mathrm{K}$). Due to the long radiative lifetime of the fine-structure
transition ($4\times10^5~\mathrm{s}$), radiative stabilization can be neglected.

Following \citet{watson1980,ponomarev1992}, who compute the autoionization rate using the formulation by \citet{seaton1976}, i. e. :
\begin{eqnarray}
A_{nl}^a=4\frac{Ry c}{h} \frac{\Omega(l)}{n^3 \omega(j,nl)},
\end{eqnarray}
with $\Omega(l)$ the collision strength for the ${^2}P_{1/2}-{^2}P_{3/2}$ excitation at the threshold. As \citet{watson1980},
we used the formula obtained by \citet{osterbrock1965}:
\begin{eqnarray}
\Omega{l}=\frac{227}{(2l-1)(2l+1)(2l+3)l(l+2)},
\end{eqnarray}
\noindent valid for $l>4$.
In order to avoid the singularity at $l=0$ we computed the autoionization rate, $A_{nl}^a$, from the approximate expression given in \citet{dickinson1981}:
\begin{eqnarray}
A_{nl}^a=2.25 \frac{2 \pi Ry c}{n^3 \left( l +1/2\right)^6},
\end{eqnarray}
\noindent which is valid for $l>10$. The dielectronic recombination rate is obtained by detailed balance:
\begin{eqnarray}
N^+_{1/2} N_e \alpha_{nl}^d= N_{nl} A_{nl}^a.
\end{eqnarray}
\citet{walmsley1982} defined $b_{di}$ as the departure coefficient when autoionization/dielectronic recombination dominate:
\begin{eqnarray}\label{eqn_bdi}
b_{di}&=& \frac{g_{1/2}N^+_{1/2}}{g_{3/2}N^+_{3/2}}  \exp\left[-\Delta E/kT_e\right],\nonumber\\
&=&\frac{1}{R}.
\end{eqnarray}

A comparison of the dielectronic recombination rate with values from the literature is hampered by the focus of previous works on higher temperatures.
The calculations of dielectronic recombination rates for Carbon from \citet{nussbaumer1983} did not include fine structure transitions and are not
suited for a direct comparison with the study presented here. Furthermore, the values presented by \citet{gu2003} are given for higher temperatures
than those studied here. The more recent study of \citet{altun2004} provides state resolved values for dielectronic recombination rates. For
the physical conditions of interest in this article and for the fine structure levels of interest here, \citet{altun2004} provides values using intermediate
coupling. However, a direct comparison with the results from \citet{altun2004} is not possible since they only include $\Delta n= 0$ type dielectronic
recombination resonances associated with the excitation of a $2s$ electron to a $2p$ level (see Equation 1 in \citealt{altun2004}).

\section{Results}\label{section_results}
The behavior of CRRLs with frequency depends on the level population of carbon via the departure coefficients. We compute departure coefficients for
carbon atoms by solving the level population equation using the rates described in Section \ref{section_rates} and the approach in Section \ref{section_nummethod}.
Here, we present values for the departure coefficients and provide a comparison with earlier studies in order to illustrate the effect of our improved rates and numerical approach.
A detailed analysis of the line strength under different physical conditions relevant for the diffuse clouds and the effects of radiative transfer are provided in
an accompanying article (Paper~II).

\subsection{Departure Coefficient for Carbon Atoms}\label{section_cmodels}
The final departure coefficients for carbon atoms (i.e. $b_n^{final}$) are obtained by computing the departure coefficients recombining from both parent ions,
those in the ${^2}P_{1/2}$ level and those in the ${^2}P_{3/2}$ level. Therefore, it is illustrative to study the individual departure coefficients for the
${^2}P_{1/2}$ core, which are hydrogenic, and the departure coefficients for the ${^2}P_{3/2}$ core separately.

\subsubsection{Departure Coefficient in the Hydrogenic Approximation}\label{section_hmodels}
\begin{figure}[!ht]
\includegraphics[width=1\columnwidth]{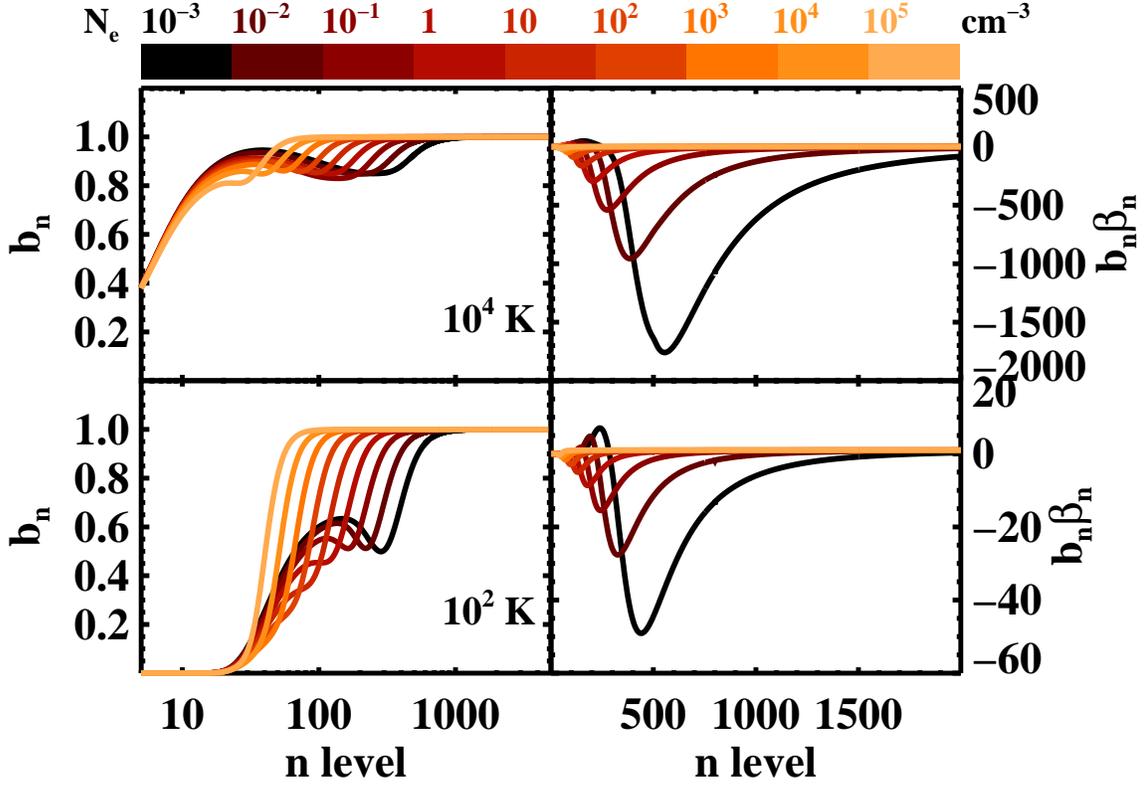}
\caption{$b_n$ values (left) and $b_n \beta_n$~values (right) for hydrogen at $T_e=10^4~\mathrm{and}~10^2~\mathrm{K}$ (upper and lower panels, respectively)
for different densities ($N_e$, colorscale). The departure coefficients obtained using the $nl$-method show a ``bump'' at low $n$~levels. The strength and
position of the ``bump'' depend on the physical conditions. As density increases, the $l$-changing collisions redistribute the electron population more effectively.\label{fig_bn1d4}}
\end{figure}

In Figure~\ref{fig_bn1d4} we show example $b_n$~and $b_n \beta_n$~values obtained in the hydrogenic approximation at $T_e=10^2~\mathrm{and}~10^4~\mathrm{K}$
for a large range in density. The behavior of the $b_n$~values as a function of $n$ can be understood in terms of the rates that are included in the
level population equation. At the highest $n$~levels, collisional ionization and three body recombination dominate the rates in the level population
equation and the $b_n$~values are close to unity. We can see that as the density increases, collisional equilibrium occurs at lower $n$ levels and
the $b_n$ values approach unity at lower levels. In contrast, for the lowest $n$ levels, the level population equation is dominated by radiative processes
and the levels drop out of collisional equilibrium. As the radiative rates increase with decreasing $n$ level, the departure coefficients become smaller.
We note that differences in the departure coefficients for the low $n$ levels for different temperatures are due to the radiative recombination rate,
which has a $T_e^{-3/2}$ dependence. 

\begin{figure}[!ht]
\includegraphics[width=0.8\columnwidth,angle=90]{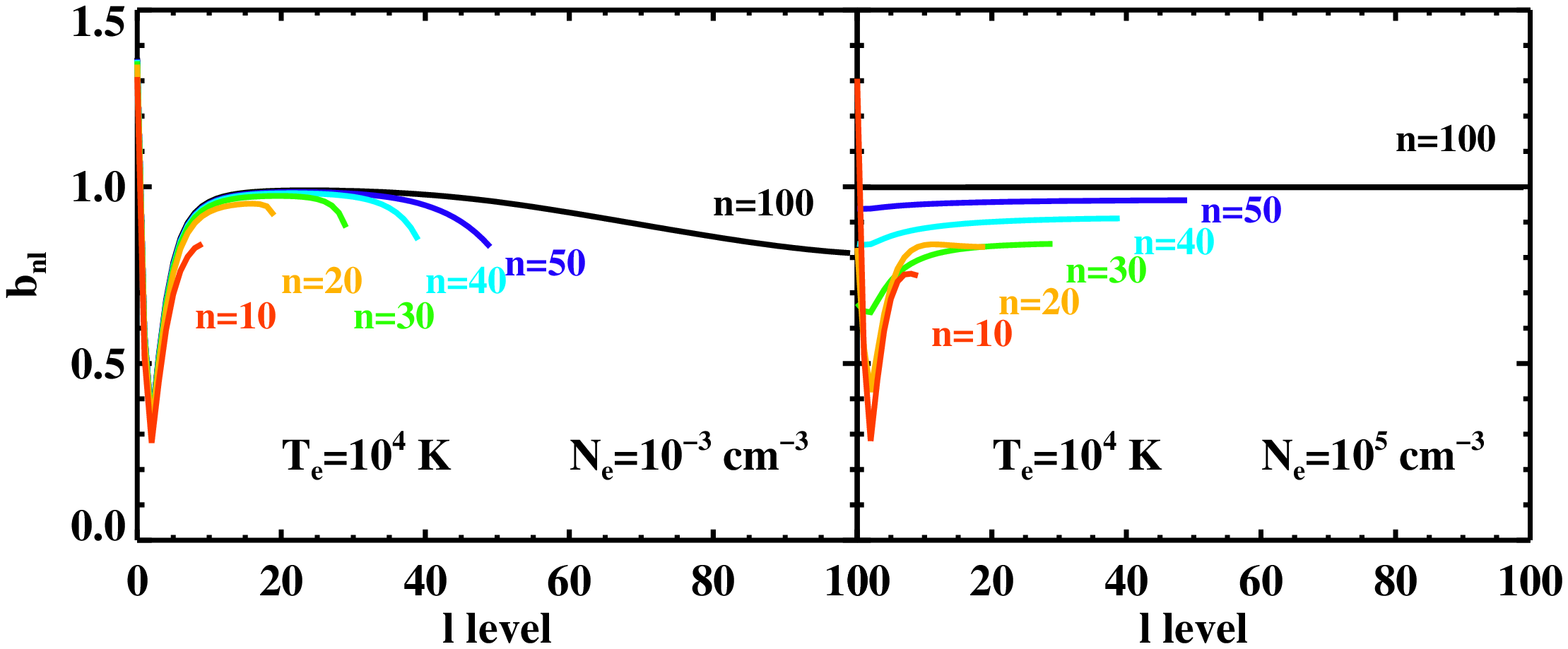}
\caption{Example of hydrogenic $b_{nl}$~values at low densities (\emph{Left panel}) and high densities (\emph{Right panel}). Statistical distribution of the $l$-sublevels
is attained at levels as low as $\sim40$. For lower levels, radiative processes dominate the level population. At low density (\emph{Left panel}), radiative processes
dominate even at high $n$~levels.\label{fig_bnl}}
\end{figure}

At intermediate $n$ levels, the behavior of the $b_n$ as a function of $n$ shows a more complex pattern with a pronounced ``bump'' in the $b_n$ values
for intermediate levels ($n\sim10~\mathrm{to}~\sim 100$). To guide the discussion we refer the reader to Figure~\ref{fig_bn1d4}. Starting at the
highest $n$, $b_n\rightarrow1$, as mentioned above. For these high $n$ levels, $l$-changing collisions efficiently redistribute the electron population
among the $l$ states and, at high density, the $b_{nl}$ departure coefficients are unity as well (Figure~\ref{fig_bnl}, upper panels). For lower values of $n$, the
$b_n$~values decrease due to an increased importance of spontaneous transitions. At these $n$ levels, the $b_n$~values obtained by the $nl$~method
differ little from the values obtained by the $n$-method, since $l$-changing collisions efficiently redistribute the electrons among the $l$ sublevels
for a given $n$ level. For lower $n$ levels, the effects of considering the $l$~sublevel distribution become important as $l$-changing
collisions compete with spontaneous decay, effectively ``storing'' electrons in high $l$ sublevels for which radiative decay is less important.
collisions compete with spontaneous decay, effectively ``storing'' electrons in high $l$ sublevels for which radiative decay is less important.
Specifically, the spontaneous rate out of a given level is approximately $A_{nl}\simeq 10^{10}/n^3/l^2 (\mathrm{s^{-1}})$, and is higher for lower $l$ sublevels.
Thus, high-$l$ sublevels are depopulated more slowly relative to lower $l$ sublevels on the same $n$ level. This results in a slight increase
in the departure coefficients. Reflecting the statistical weight factor in Equation~\ref{eqn_bn2}, the higher $l$ sublevels dominate the final $b_n$~value
resulting in an increase in the final $b_n$~value. As the density increases, the $l$ sublevels approach statistical distribution faster. As a result,
the influence of the $l$ sublevel population on the final $b_n$~is larger for lower densities than for higher densities at a given $T_e$.
The interplay of the rates produce the ``bump'' which is apparent in the $b_n$ distribution (Figure \ref{fig_bn1d4})

The influence of $l$-changing collisions on the level populations and the resulting increase in the $b_n$ values was already presented by \citet{hummer1987} and
analyzed in detail by \citet{strelnitski1996} in the context of hydrogen masers. The results of our level population models are in good agreement with those
provided by \citet{hummer1987} as we show in Section \ref{section_comparison}.

\subsubsection{Departure Coefficient for Carbon Atoms Including Dielectronic Recombination}
Only carbon atoms recombining to the ${^2}P_{3/2}$ ion core are affected by dielectronic recombination. Having analyzed the departure coefficients
for the hydrogenic case, we focus now on the $b_{nl}^{3/2}$ values and the resulting $b_n^{final}$ as introduced in Section~\ref{section_levelpop}.

\begin{figure}[!ht]
\includegraphics[width=1\columnwidth]{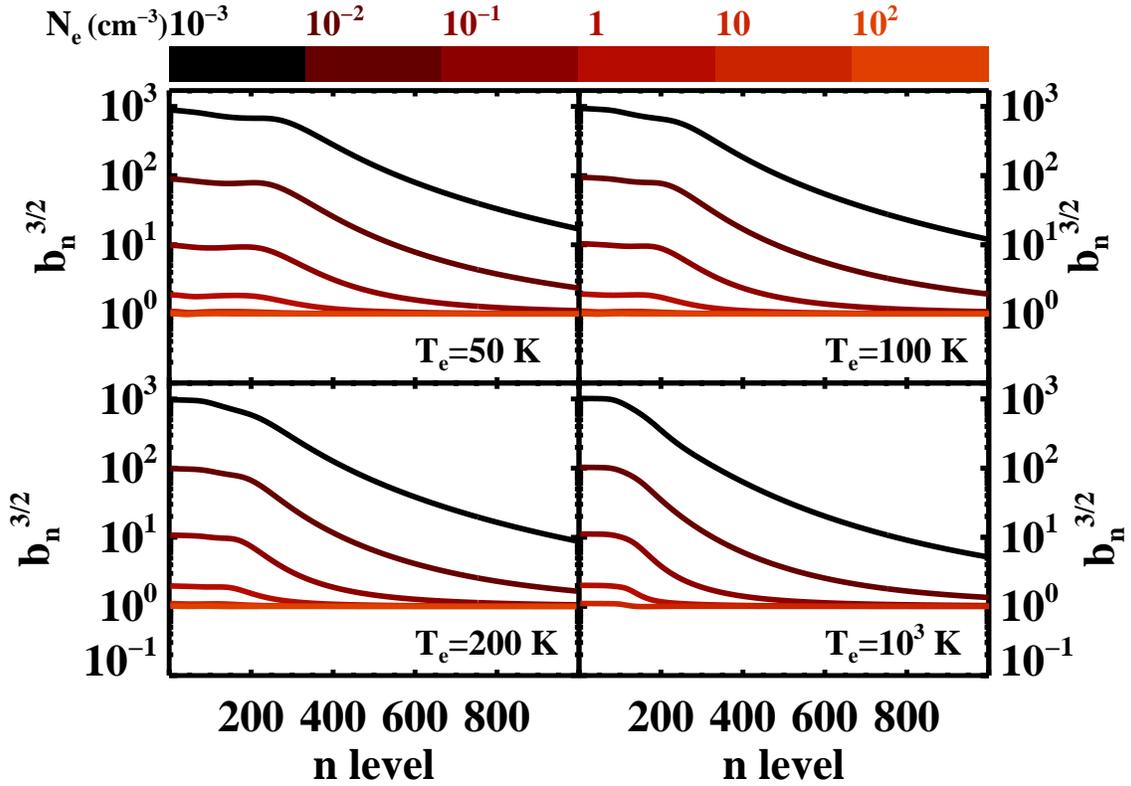}
\caption{Departure coefficients for the ${^2}P_{3/2}$ parent ions as a function of $n$ at $T_e=50,~100,~200~\mathrm{and}~1000~\mathrm{K}$ for different densities ($N_e$, colorscale).
The values for low $n$ levels are close to $b_{di}$ and decrease towards a value of one. At high densities, $b_n^{3/2}\approx 1$.\label{fig_bn32carbon}}
\end{figure}

\begin{figure}[!ht]
\includegraphics[width=1\columnwidth]{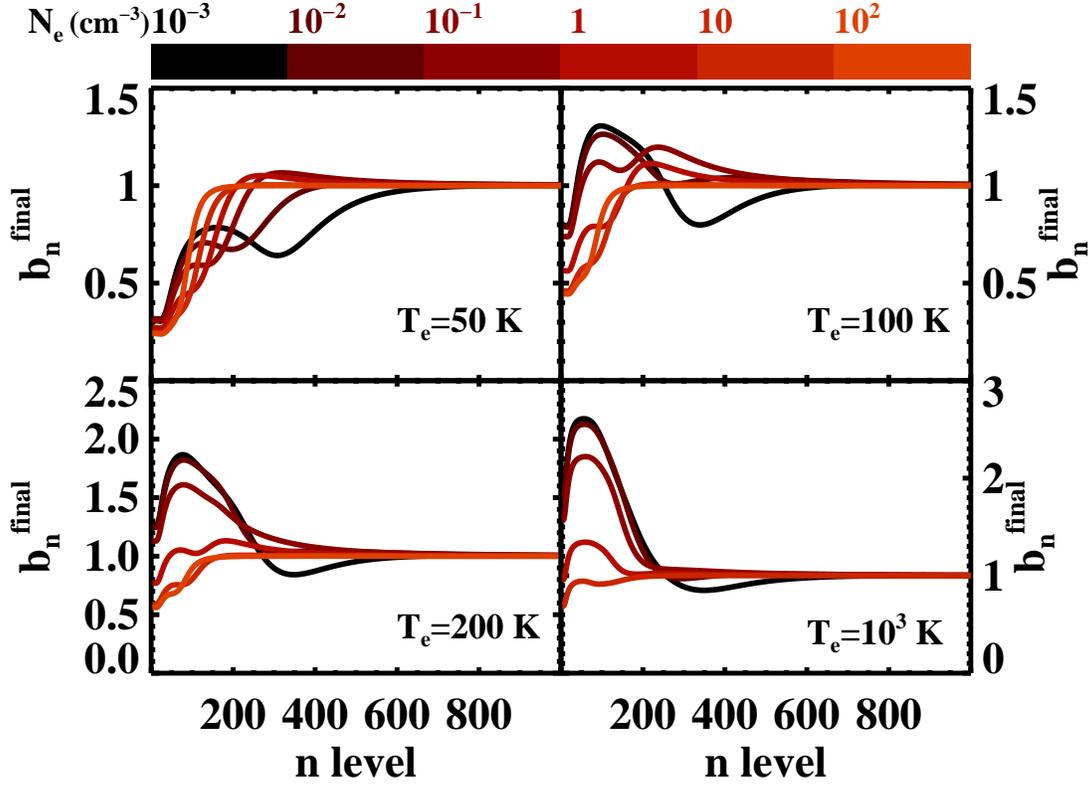}
\caption{Final departure coefficients for carbon atoms ($b_n^{final}$) as a function of $n$ level at  $T_e=50,~100,~200~\mathrm{and}~1000~\mathrm{K}$ for different densities ($N_e$, colorscale).
The ``bump'' seen in hydrogenic atoms is amplified by dielectronic recombination. As density increases the $b_n^{final}$ are closer to the hydrogenic value.\label{fig_bnfinalcarbon}}
\end{figure}

Figure~\ref{fig_bn32carbon} show example values for $b_n^{3/2}$ for $T_e=50,~100,~200~\mathrm{and}~1000~K$ and electron densities between $10^{-3}~\mathrm{and}~10^2~\mathrm{cm^{-3}}$.
As pointed out by \citet{watson1980}, the low lying $l$ sublevels are dominated by the dielectronic process and the $b_{nl}^{3/2}$ values are equal to $b_{di}$ (Equation~\ref{eqn_bdi}).
As can be seen in Figure~\ref{fig_rvalue}, such values can be much larger than unity at low densities resulting in an overpopulation of the low $n$ levels for the $3/2$
ion cores. In Figure~\ref{fig_bnfinalcarbon} we show $b_n^{final}$ as a function of $n$ level under the same conditions. We see that at high electron densities the departure
coefficients show a similar behavior as the hydrogenic values. Furthermore, an increase in the level population to values larger than unity is seen at low densities and
moderate to high temperatures.

To guide the discussion, we analyze the behavior of the $b_n^{final}$ when autoionization/dielectronic recombination dominates. This occurs at different levels
depending on the values of $T_e$ and $N_e$ considered. Nevertheless, it is instructive to understand the behavior of the level population in extreme cases.
When autoionization/dielectronic recombination dominate, the $b_n^{final}$ in Equation~\ref{equation_bnfinal2} is given by:
\begin{eqnarray}\label{equation_bnfinal3}
b_n^{final} &\approx& \frac{b_n^{1/2}+\left[N^+_{3/2}/N^+_{1/2}\right]_{LTE}}{1+R \left[N^+_{3/2}/N^+_{1/2}\right]_{LTE}}.
\end{eqnarray}
\noindent At high densities, $R$ approaches unity and we note two cases. The first case is when $T_e$ is high, the maximum value of $\left[N^+_{3/2}/N^+_{1/2}\right]_{LTE}=2$,
meaning that a large fraction of the ions are in the ${^2}P_{3/2}$ core. Consequently, $b_n^{final} \approx (b_n^{1/2}+2)/3$, thus the effect of dielectronic recombination
is to increase the level population as compared to the hydrogenic case. We also note that since $b_n^{1/2} \leq 1$ the final $b_n^{final} \leq 1$.
The second case we analyze is for low $T_e$, where the ion LTE ratio is low and most of the ions are in the ${^2}P_{1/2}$ core.
Thus, $b_n^{final} \approx b_n^{1/2}$ and the departure coefficients are close to hydrogenic.

At low densities, $R\ll 1$ and, as above, we study two cases. The first is when $T_e$ is high, the maximum value of $\left[N^+_{3/2}/N^+_{1/2}\right]_{LTE}=2$ and
$b_n^{final} \approx b_n^{1/2}+2$, therefore dielectronic recombination produces a large overpopulation as compared to the hydrogenic case. The second case is when
$T_e$ is low and most of the ions are in the ${^2}P_{1/2}$ level and, as in the high density case, the $b_n^{final} \approx b_n^{1/2}$. We note from this analysis
that overpopulation of the $b_n^{final}$ (relative to the hydrogenic case) is only possible for a range of temperatures and densities. In particular,
$b_n^{final}$ is maximum for high temperatures and low densities.

Having analyzed the behavior of the $b_n^{final}$ values in the extreme $b_n^{3/2}=b_{di}$ case, now we analyze the behavior of $b_n^{3/2}$ with $n$. The population
in the low $n$ levels is dominated by dielectronic recombination \citep{watson1980,walmsley1982} and $b_n^{3/2}=b_{di}$ up until a certain $n$ level where $b_n^{3/2}$
begins to decrease down to a value of one. The $n$ value where this change happens depends on temperature, moving to higher $n$ levels as $T_e$ decreases.
To understand this further, we analyze the rates involved in the $l$ sublevel population (Figure \ref{fig_bn32carbon}). The low $l$ sublevels are dominated by dielectronic
recombination and autoionization and the $b_{nl}$ values for the $3/2$ ion cores are $b_{nl}^{3/2}=b_{di}$.
For the higher $l$ sublevels other processes (mainly collisions) populate or depopulate electrons from the level $n$ and the net rate is lower than that of
the low $l$ dielectronic recombination/autoionization. This lowers the $b_{nl}$ value, which is effectively delayed by $l$-changing collisions since they redistribute the
population of electrons in the $n$ level. The $b_{nl}$ for highest $l$ values dominate the value of $b_n^{3/2}$ due to the statistical weight factor.

\begin{figure}[!ht]
\includegraphics[width=1\columnwidth]{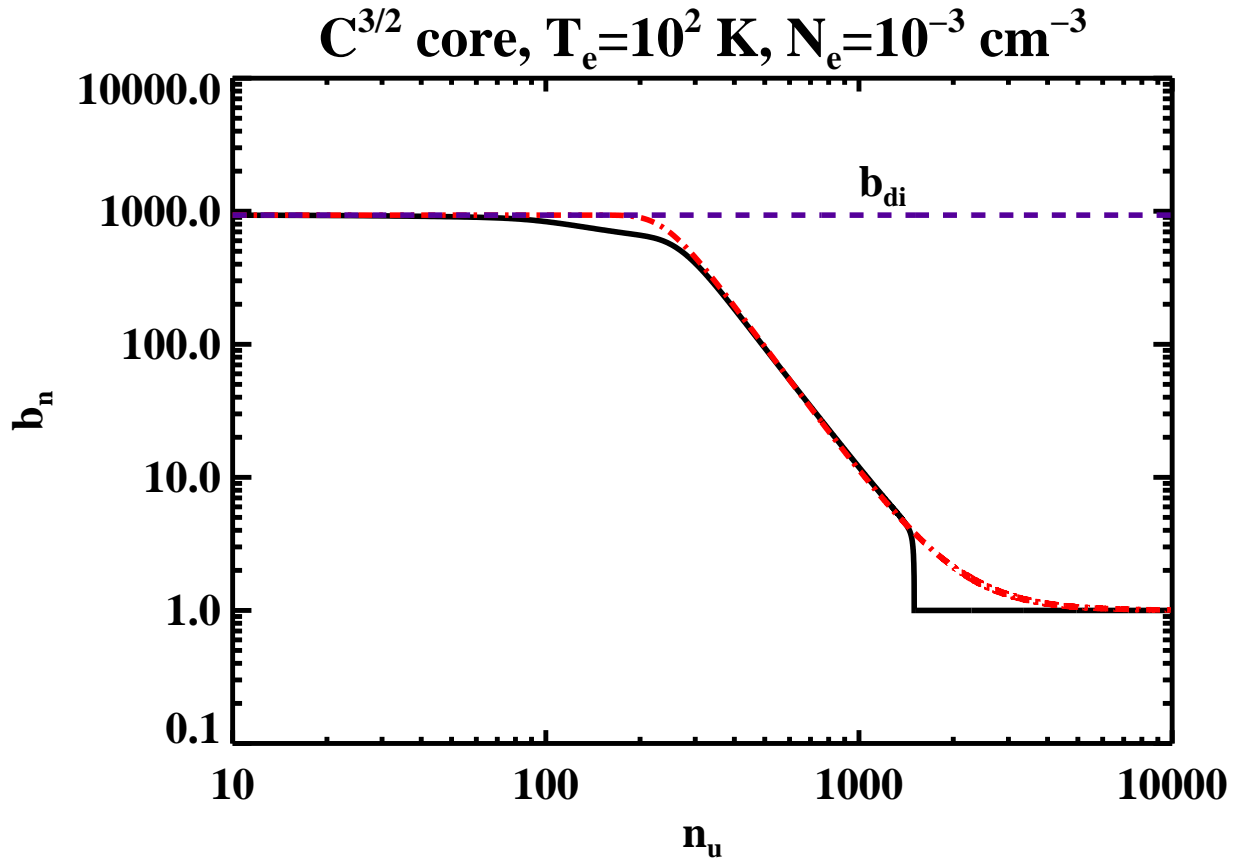}
\caption{$b_n^{3/2}$~values for carbon as a black line (solid), the discontinuity at $n=1500$ is due to the $n_{crit}$ value.
Overplotted as a red (dot dashed) line is the approximation in Equation~\ref{eqn_appbn32}. The blue (dashed) line is the
value of $b_{di}$.\label{fig_exbn32}}
\end{figure}

We note that the behavior of the $b_n^{3/2}$~cores as a function of $n$ (see Figure~\ref{fig_exbn32}) can be approximated by:
\begin{eqnarray}\label{eqn_appbn32}
b_n^{3/2}\approx \tanh \left( \left[ \frac{l_m}{n} \right]^3 \right)\times \left(b_{di}-1\right) +1.
\end{eqnarray}
\noindent with $b_{di}$~ defined as in \citet{walmsley1982} (Equation~\ref{eqn_bdi}) and $l_m$ was derived from fitting our results:
\begin{eqnarray}\label{eqn_lm}
l_m\approx 60\times\left(\frac{N_e}{10}\right)^{-0.02}\left(\frac{T_e}{10^4}\right)^{-0.25}
\end{eqnarray}

\begin{figure}[!ht]
\includegraphics[width=1\columnwidth]{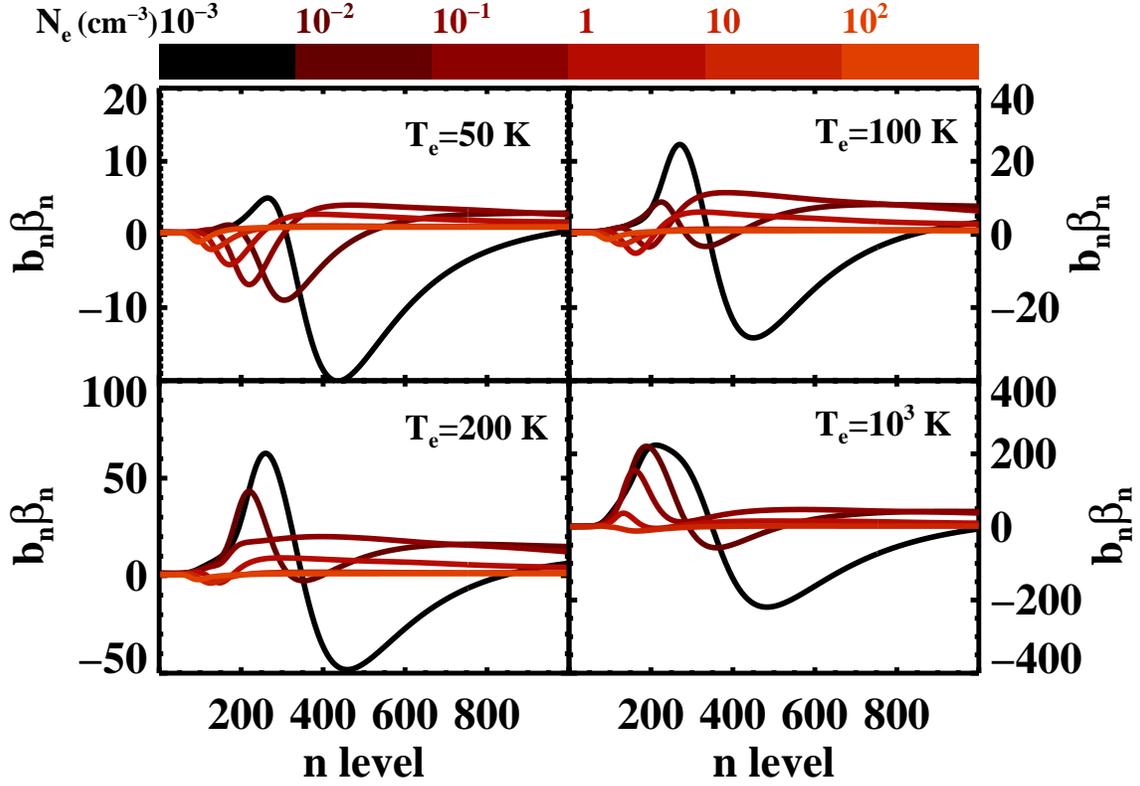}
\caption{$b_n \beta_n$ values for carbon atoms at  $T_e=50,~100,~200~\mathrm{and}~1000~\mathrm{K}$ for different densities ($N_e$, colorscale).\label{fig_bnbetacarbon}}
\end{figure}

\begin{figure}[!ht]
\includegraphics[width=1\columnwidth]{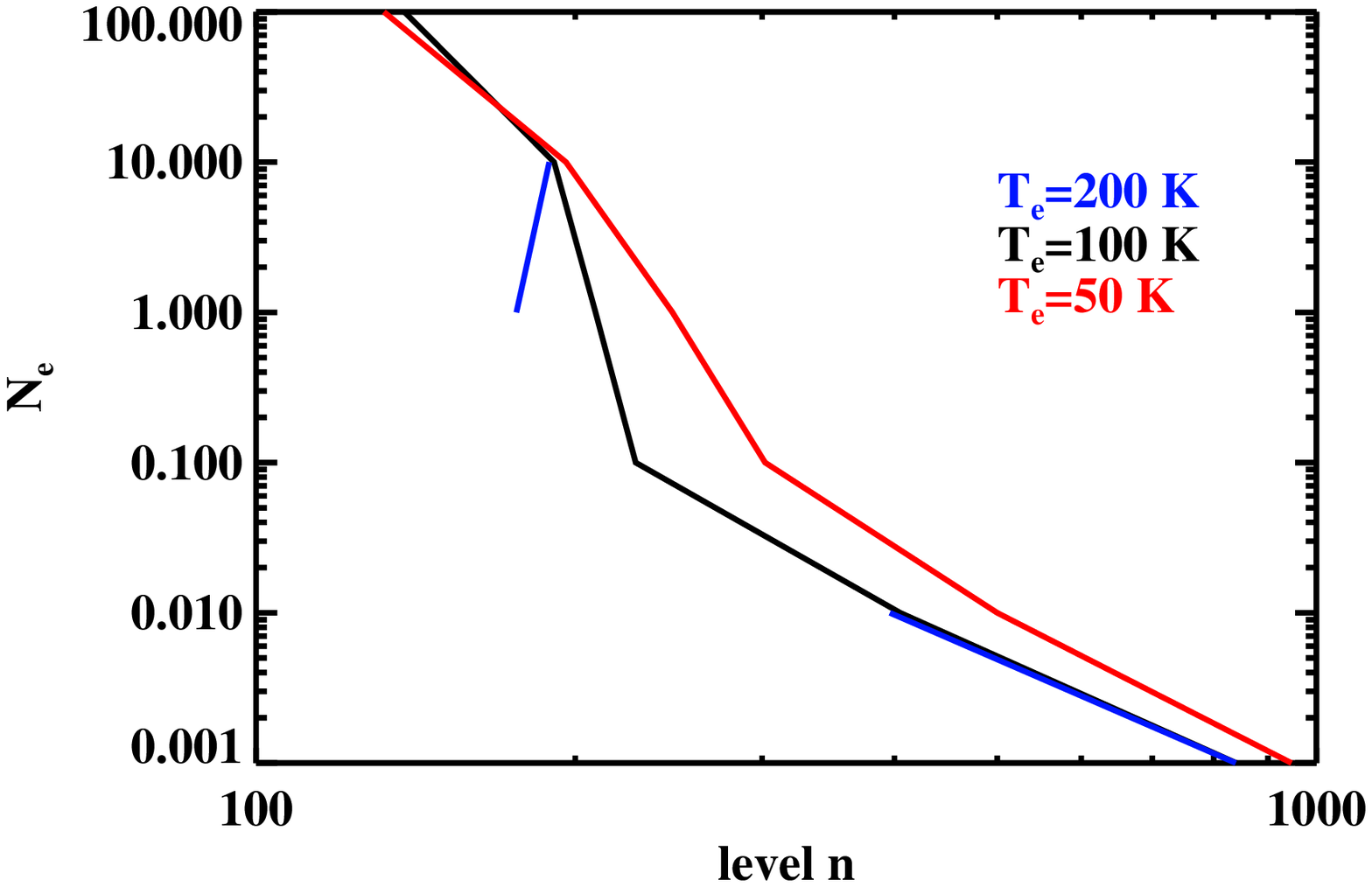}
\caption{Level where the $b_n\beta_n$ values go to zero for $T_e=50,~100~\mathrm{and}~200~\mathrm{K}$. At temperatures larger than
200 K and for an electron densities around $10^{-1}~\mathrm{cm^{-3}}$ the $b_n\beta_n$ values do not go through zero.\label{fig_bnbetanzero}}
\end{figure}

In diffuse clouds the integrated line to continuum ratio is proportional to $b_n \beta_n$. We note that the $\beta_n$ behavior is more complex as can be seen in Figure~\ref{fig_bnbetacarbon}. The low $n$ ``bump'' on the $b_n^{final}$ makes the $b_n \beta_n$ high at
low densities and for levels between about $150$ and $300$. Since the $b_n^{final}$ values decrease from values larger than one to approximately one, the 
$\beta_n$ changes sign. In Figure~\ref{fig_bnbetanzero} we show the electron density as a function of the level where the change of sign on the $b_n \beta_n$ occurs.
At temperatures higher than about $200$, our models for $N_e=0.1~\mathrm{cm^{-3}}$ show no change of sign due to the combined effects of $l$-changing collisions
and dielectronic recombination.

\subsection{Comparison with Previous Models}\label{section_comparison}
The level population of hydrogenic atoms is a well studied problem. Here, we will describe the effects of the updated collision rates as well as point out differences
due to the improved numerical method.

\subsubsection{Hydrogenic Atoms}
\begin{figure}[!ht]
\includegraphics[width=0.5\columnwidth]{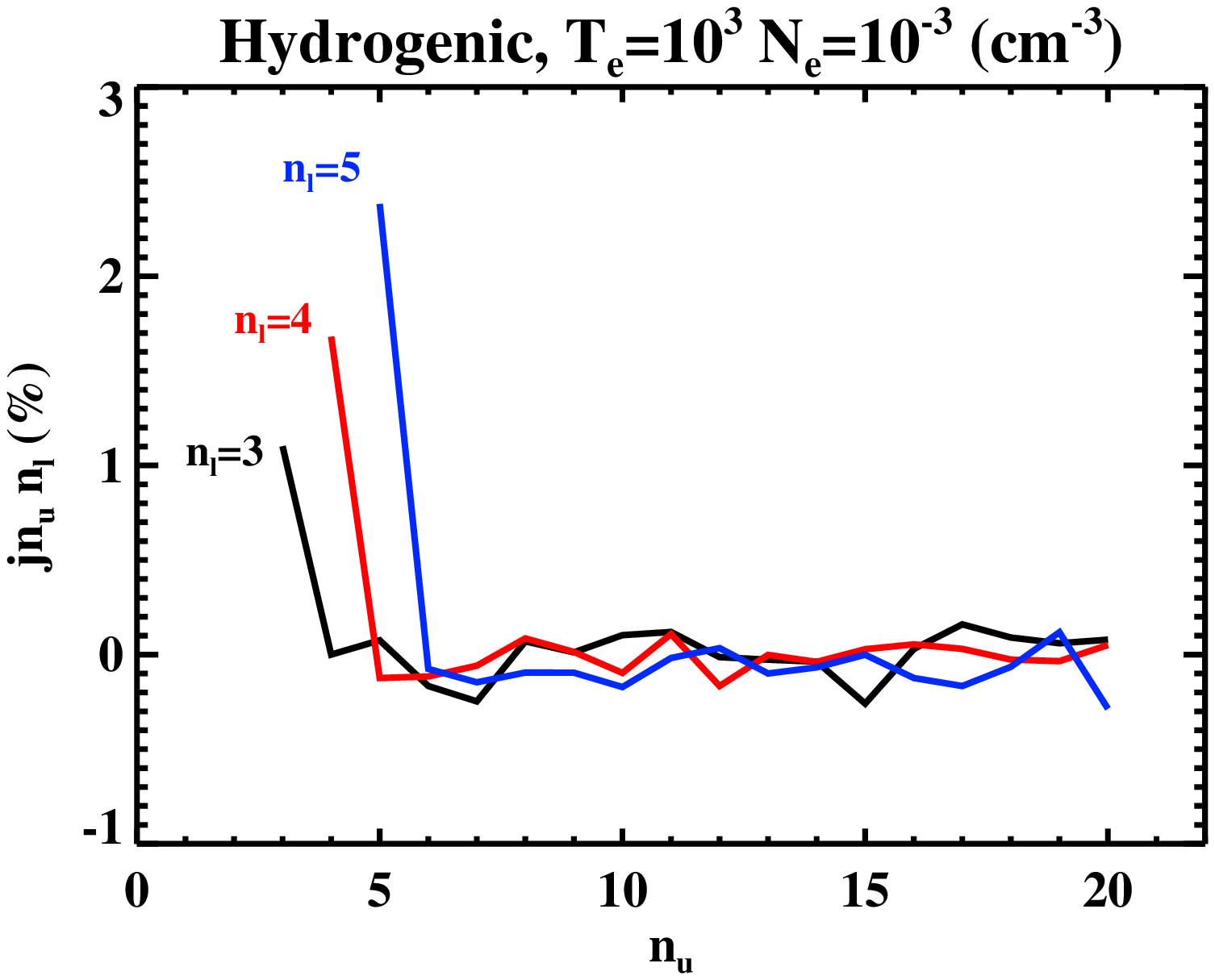}
\includegraphics[width=0.5\columnwidth]{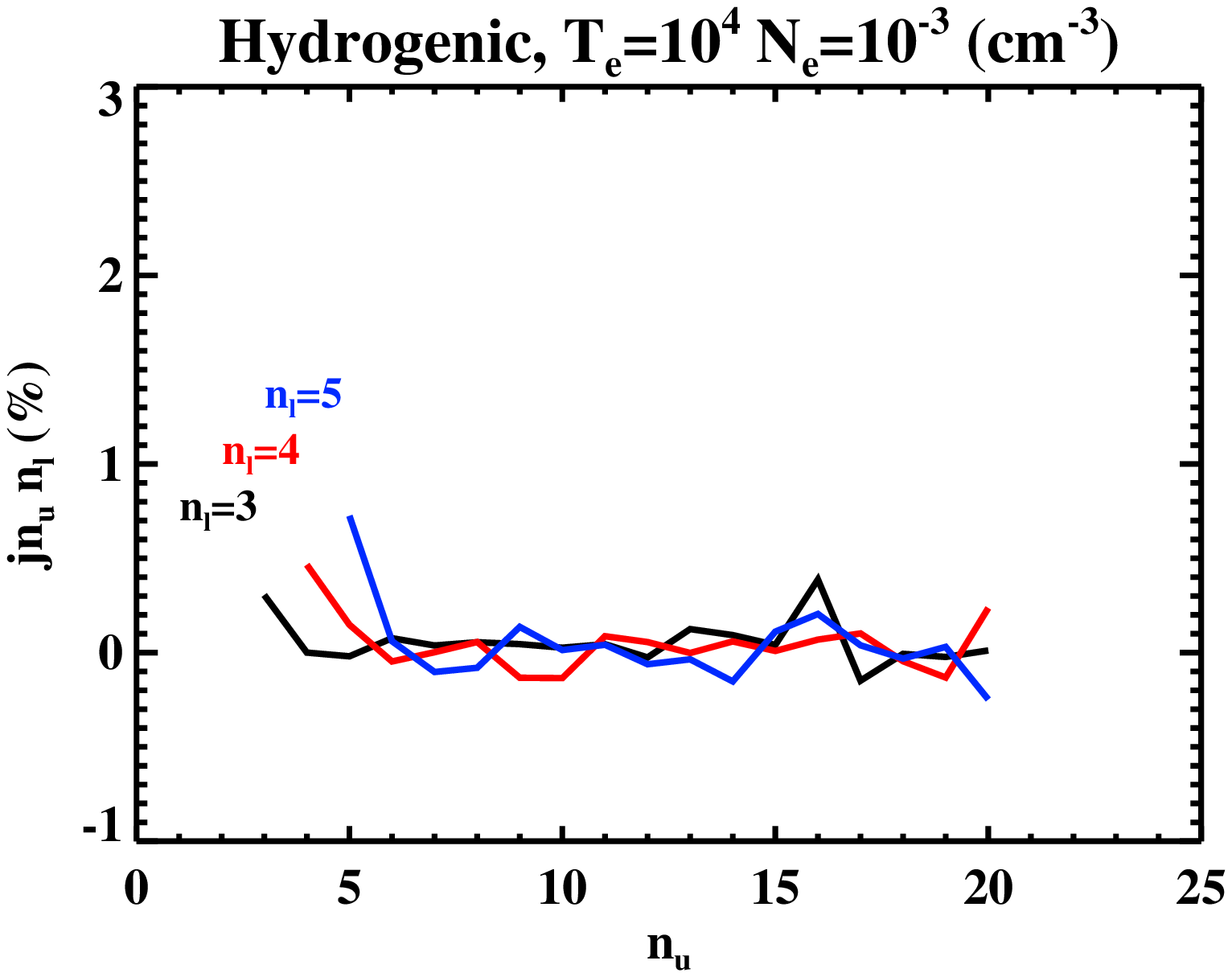}
\caption{Difference between the emissivities (normalized to H$\beta$) for low $nl$~lines at low density and the results from \citet{martin1988} in the low $N_e$~approximation.
Our results agree to better than 1\% at most levels.\label{fig_compmartin}}
\end{figure}

At the lowest densities, we can compare our results for hydrogenic atoms with the values of \citet{martin1988} for Hydrogen atoms. The results of \citet{martin1988} were
obtained in the low density limit, i.e. no collision processes were taken into account in his computations. The results are given in terms of the emissivity of the line
normalized by the H$\beta$~emissivity. As can be seen in Figure \ref{fig_compmartin}, our results agree to better than 5\%, and for most levels to better than 0.5\%. 

\begin{figure}[!ht]
\includegraphics[width=0.5\columnwidth]{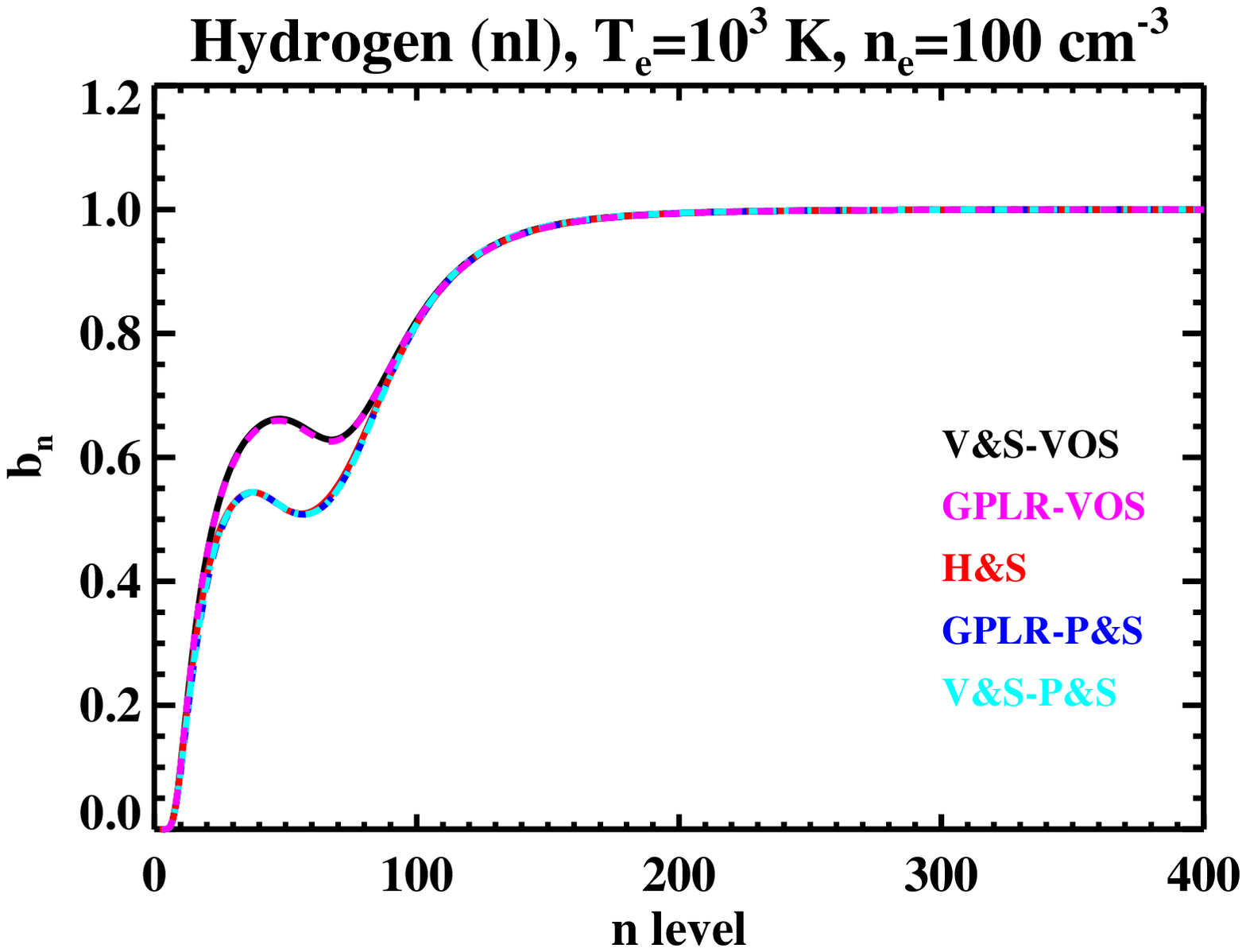}
\includegraphics[width=0.5\columnwidth]{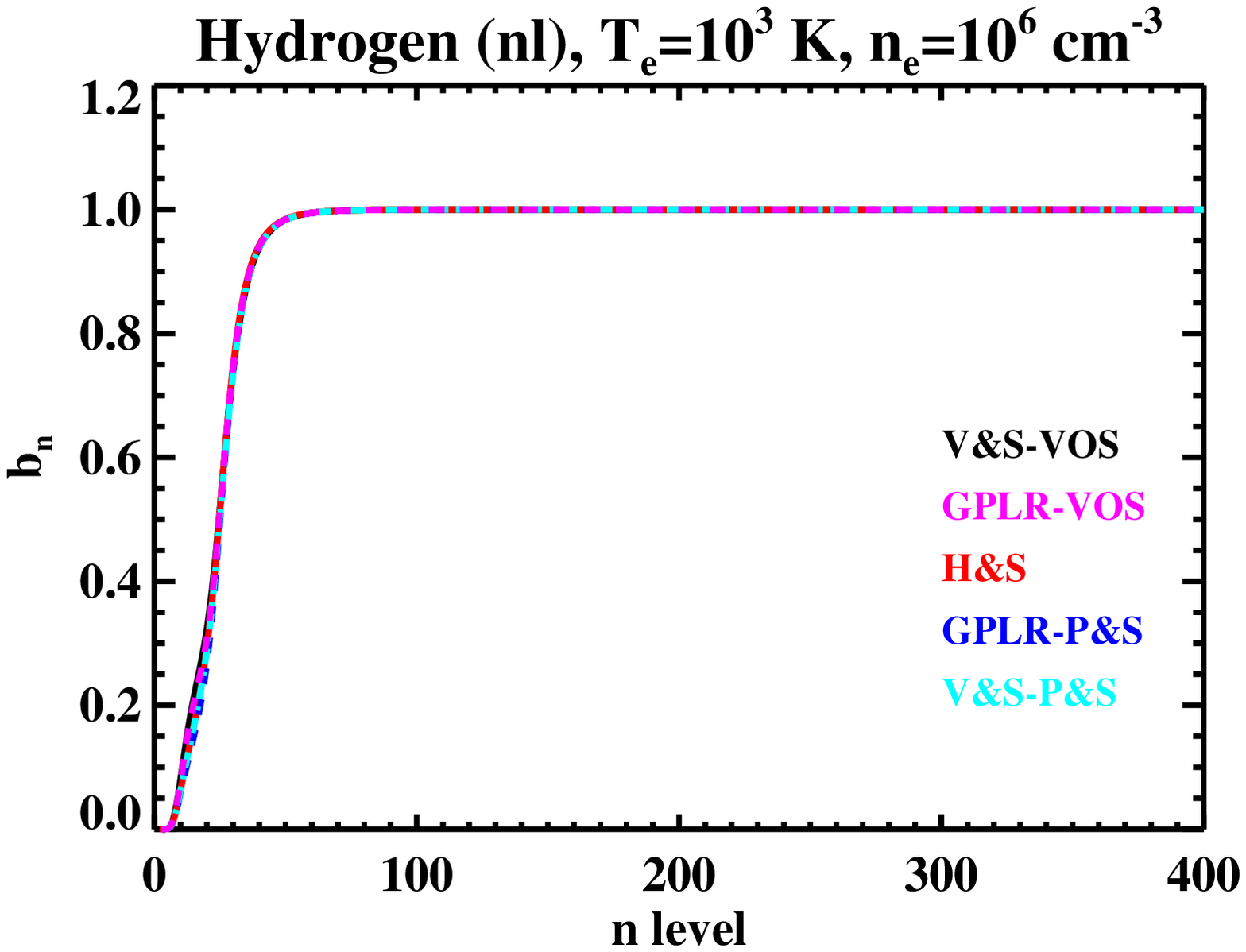}
\caption{A comparison of the effect of different collision rates on the final $b_n$~values for $T_e=1000~\mathrm{K}$~and $N_e=100~\mathrm{and}~10^6~\mathrm{cm^{-3}}$.
H\&S are the departure coefficients from \citet{hummer1987} who used \citet{gee1976}; GPLR corresponds to the use of $n$ changing collision rates from \citet{gee1976},
V\&S from \citet{vriens1980}; P\&S corresponds to the use of $l$-changing collision rates from \citet{pengellyandseaton1964},  VOS corresponds to \citet{vrinceanu2012}.
The largest differences are $\sim 30\%$ due to the use of different $l$-changing collision rates.\label{fig_effofcol}}
\end{figure}

\begin{figure}[!ht]
\includegraphics[width=1\columnwidth]{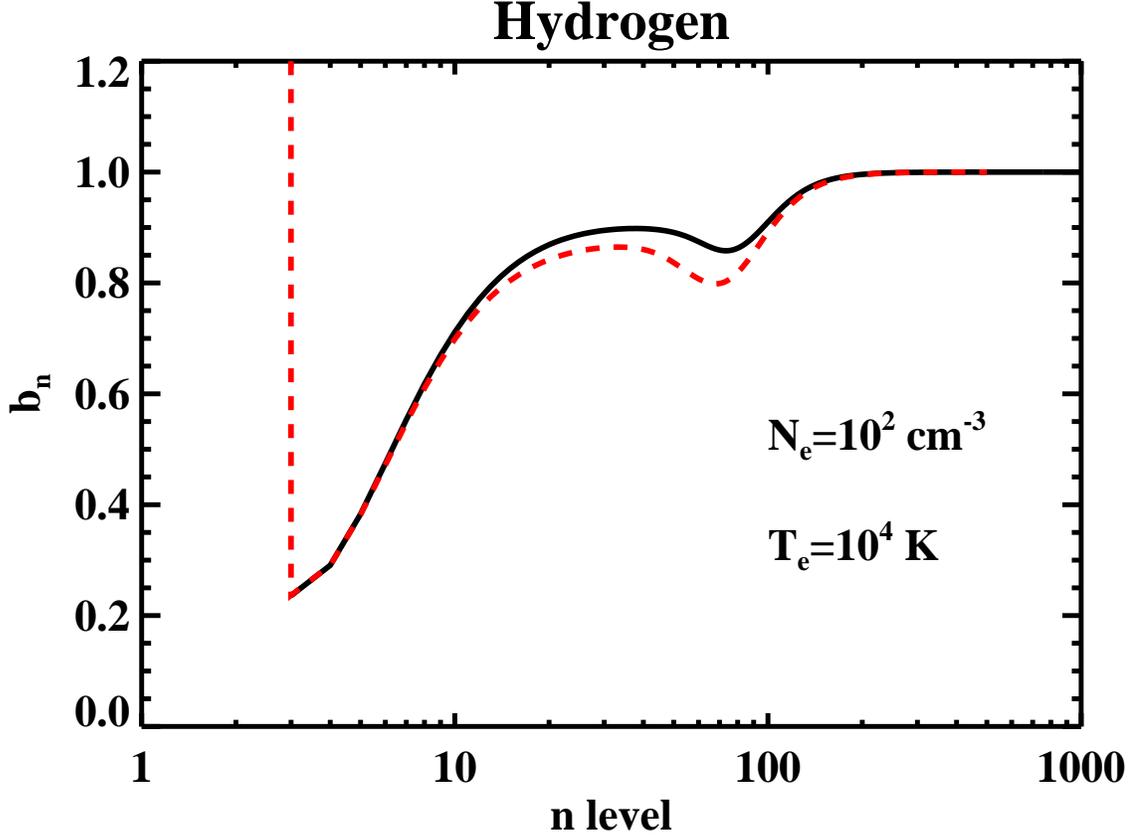}
\caption{Comparison between our $b_n$~values (black) at $T_e=10^4~\mathrm{K}$~and $N_e=100~\mathrm{cm^{-3}}$~and the results from \citet{hummer1992} (red line, dashed).
Differences are due to the use of $l$-changing collision rates from \citet{vrinceanu2012}.\label{fig_comphs}}
\end{figure}

At high densities, we compare the hydrogenic results obtained here with those of \citet{hummer1987}. Our approach reproduces well the $b_{nl}$~(and $b_n$) values of \citet{hummer1987}
(to better than 1\%) when using the same collision rates \citep{gee1976,pengellyandseaton1964} as can be seen in Figure~\ref{fig_effofcol}. We note that the effect of using
different energy changing rates ($C_{n,n'}$) has virtually no effect on the final $b_{n}$~values. On the other hand, using \citet{vrinceanu2012} values for
the $C_{nl,nl\pm1}$~rates results in differences in the $b_n$~values of 30\% at $T_e=10^3~\mathrm{K},~N_e=100~\mathrm{cm^{-3}}$. As expected, the difference
is less at higher temperatures and densities since values are closer to equilibrium (see Figure~\ref{fig_comphs}). At low $n$ levels, our results for high $l$~levels
are overpopulated as compared to the values of \citet{hummer1987} leading to an increases in the $b_n$~values.

\subsubsection{Carbon}

\begin{figure}[!ht]
 \includegraphics[scale=1,bb=14 14 133 122, width=0.313\columnwidth,angle=-0.8, trim=0 -0.12cm 0 0]{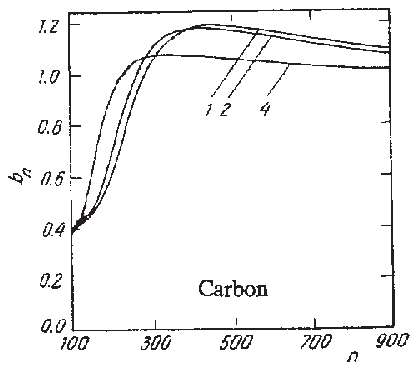}
 \includegraphics[trim=0 0 110 0,clip=true,width=0.33\columnwidth]{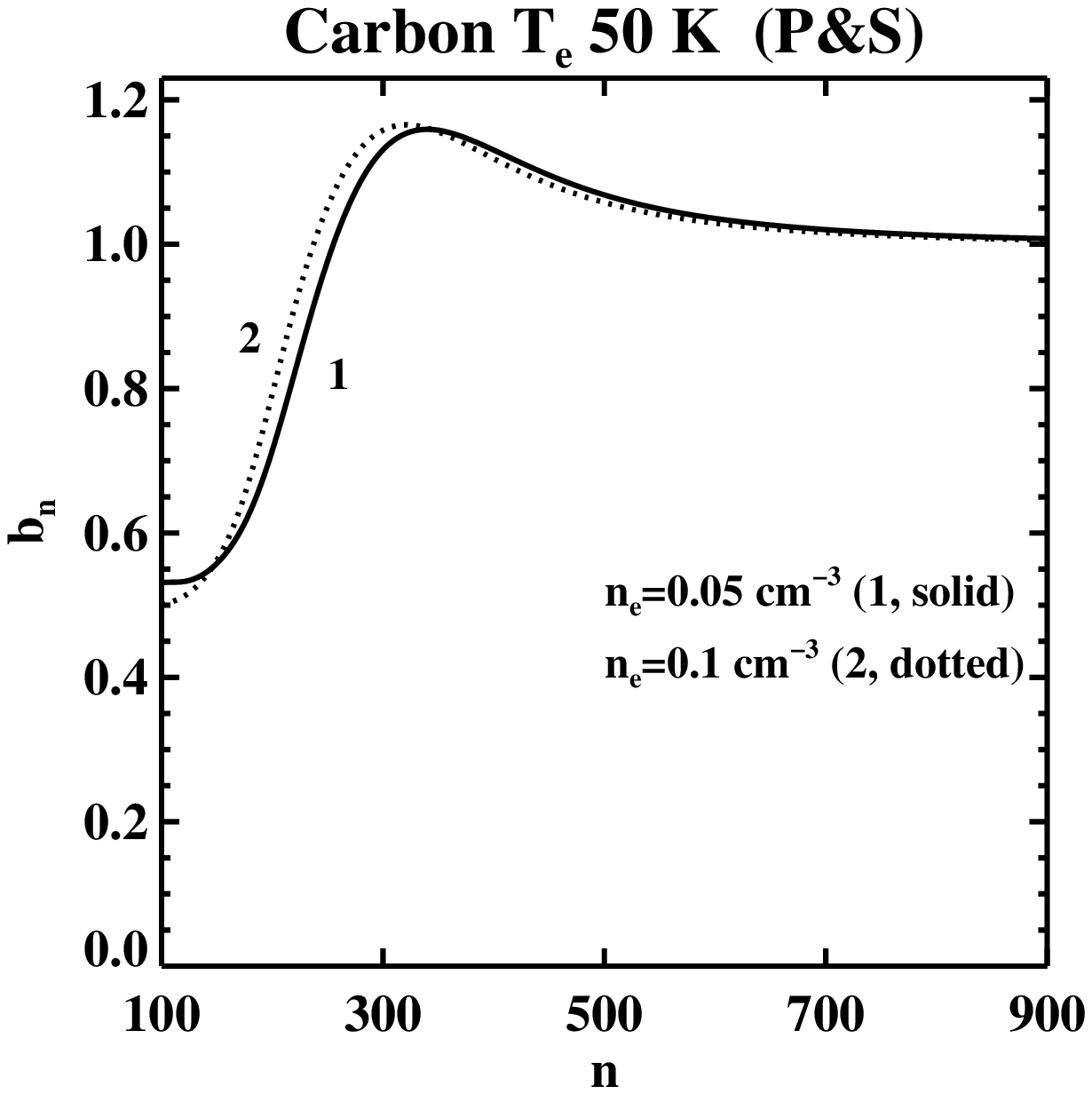}
 \includegraphics[trim=0 0 110 0,clip=true,width=0.33\columnwidth]{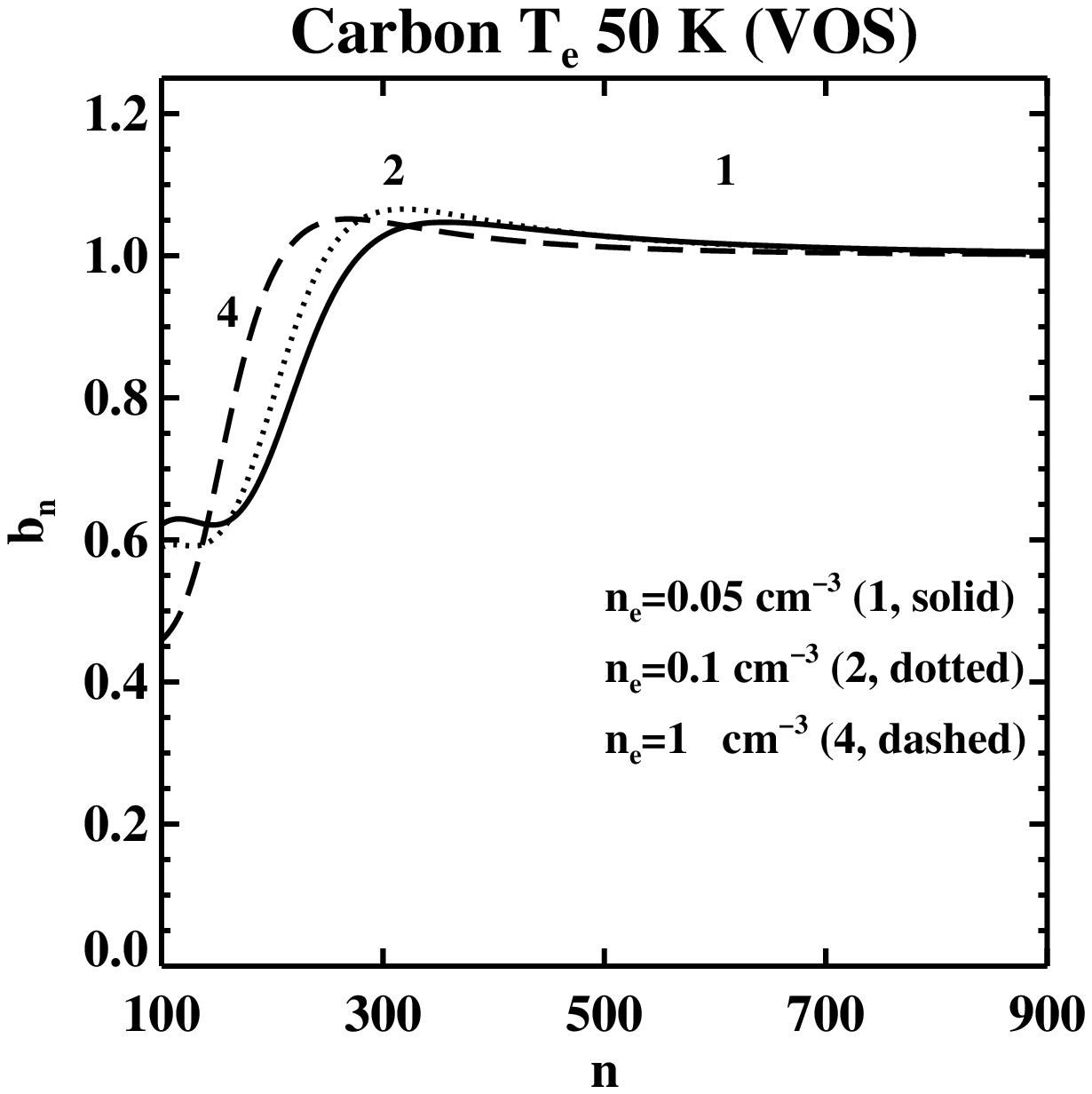}
 \includegraphics[scale=1,bb=14 14 140 122,width=0.323\columnwidth,angle=-0.9, trim=0 -0.12cm 0 0]{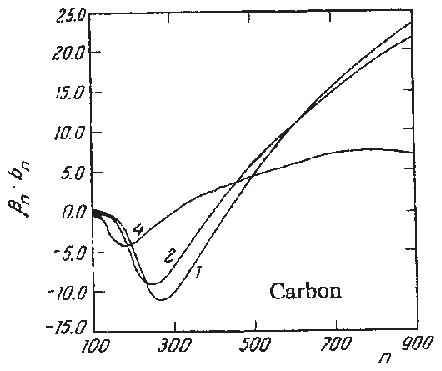}
 \includegraphics[trim=0 0.1cm 110 0,clip=true,width=0.33\columnwidth]{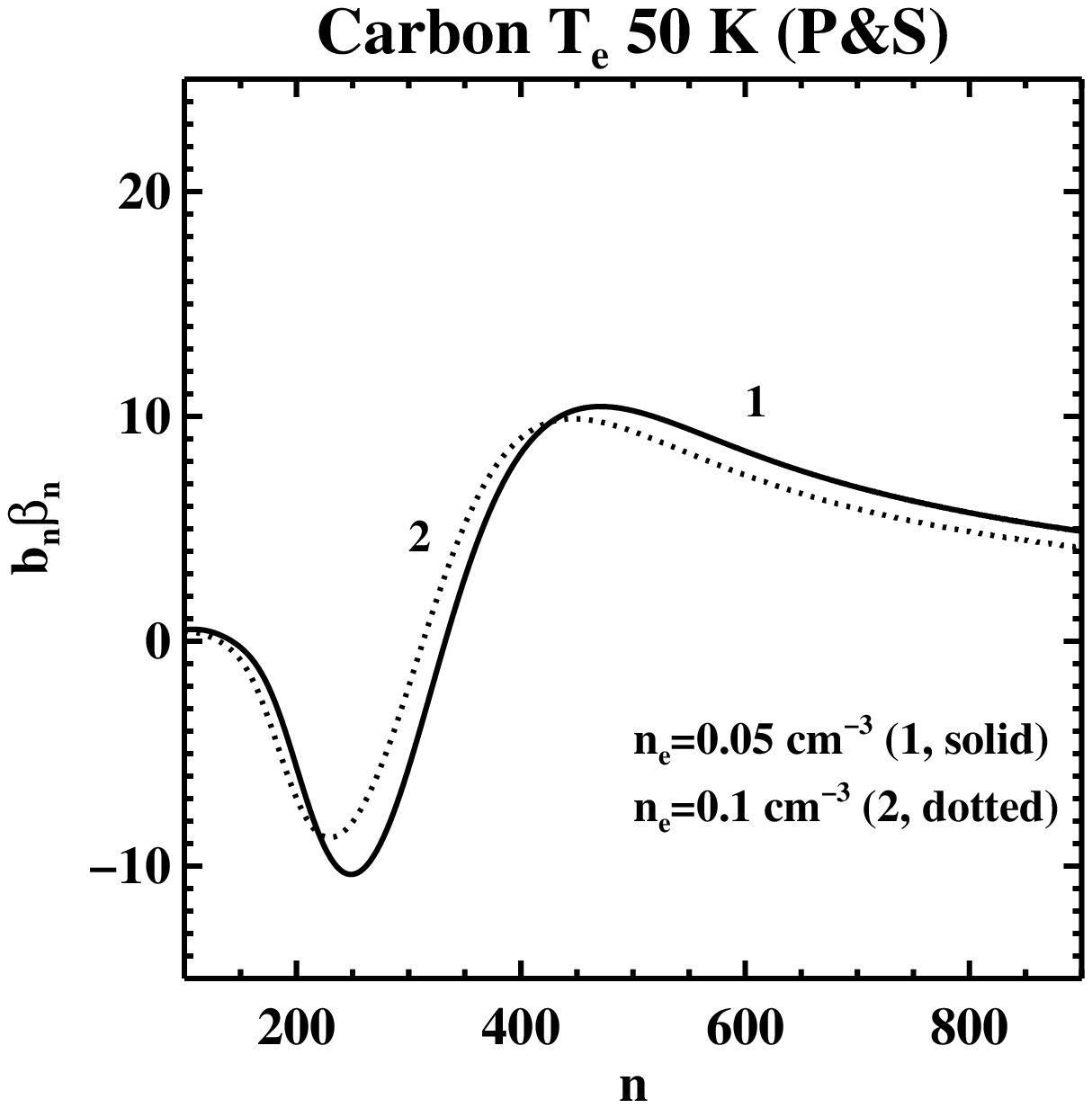}
 \includegraphics[trim=0 0 110 0,clip=true,width=0.33\columnwidth]{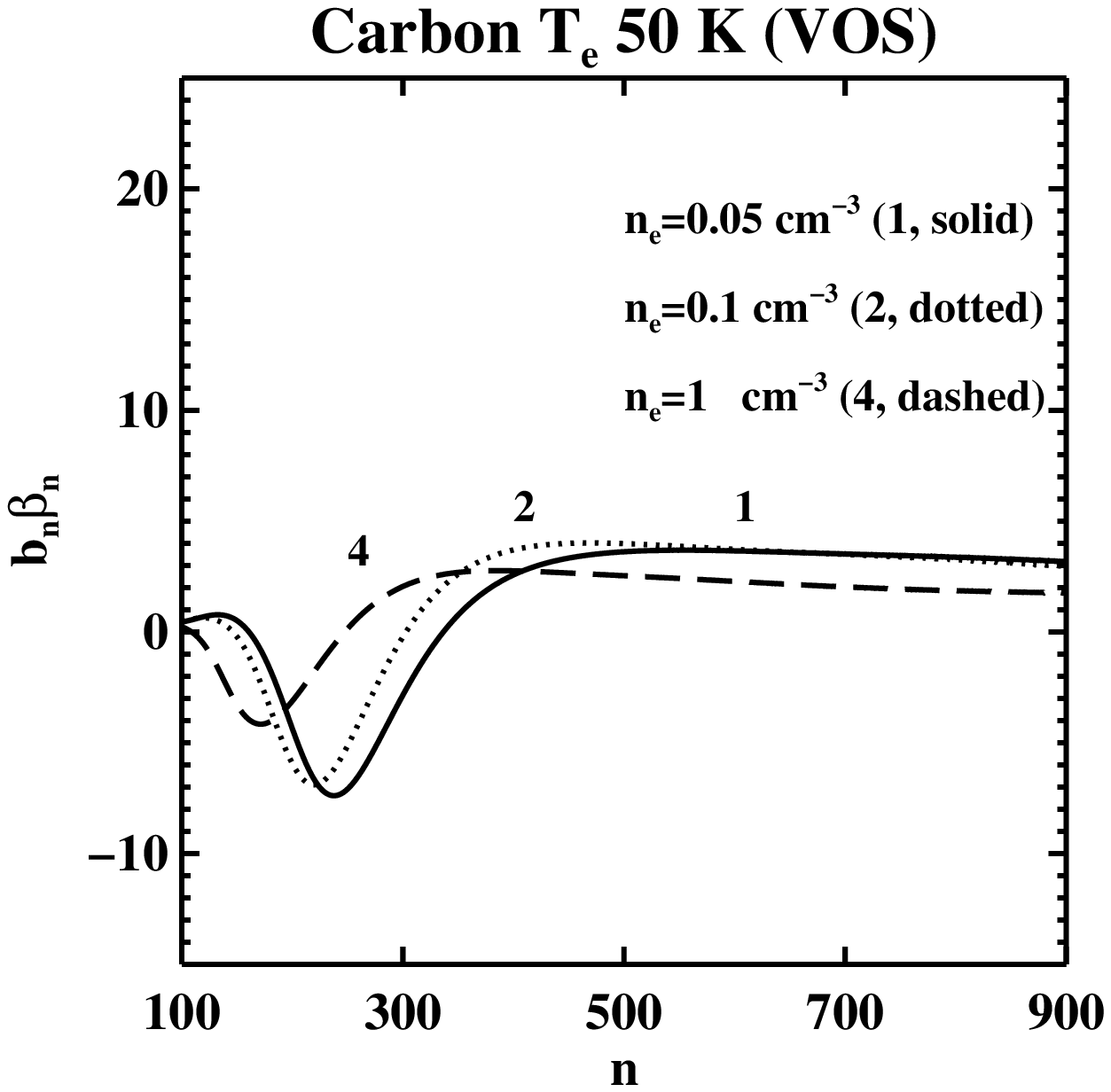}
\begin{picture}(10,10)
\put( 36,270){a)}
\put(186,270){b)}
\put(346,270){c)}
\put( 36,125){d)}
\put(186,125){e)}
\put(346,125){f)}
\end{picture}
\caption{Comparison between the CRRL departure coefficients from \citet{ponomarev1992}
(Panels a) and d); reproduced with permission from Ponomarev V.O. \& Sorochenko R. L., 1992, Soviet Astronomy Letters, 18, 215. Copyright 1992, AIP Publishing LLC.), this work using $l$-changing collision rates from \citet{pengellyandseaton1964} (Panels b) and e)) and those from \citet{vrinceanu2012} (Panels c) and f)) at $T_e$=50~K. Lines marked as 1, 2, 4 correspond to electron densities $n_e=$0.05, 0.1 and 1.0~cm$^{-3}$ (solid, dotted and dashed lines) respectively. The top panels show $b_n$ vs. $n$ and the bottom panels show the product $b_n\beta_n$ vs. $n$.\label{fig_compps}}
\end{figure}

\begin{figure}[!ht]
\includegraphics[scale=1,bb=14 14 133 121,width=0.313\columnwidth,angle=-0.5, trim=0 -0.12cm 0 0]{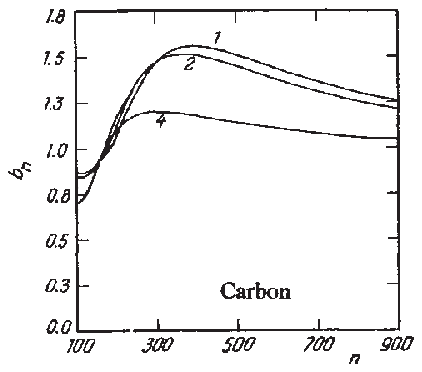}
\includegraphics[trim=0 0 110 0,clip=true,width=0.33\columnwidth]{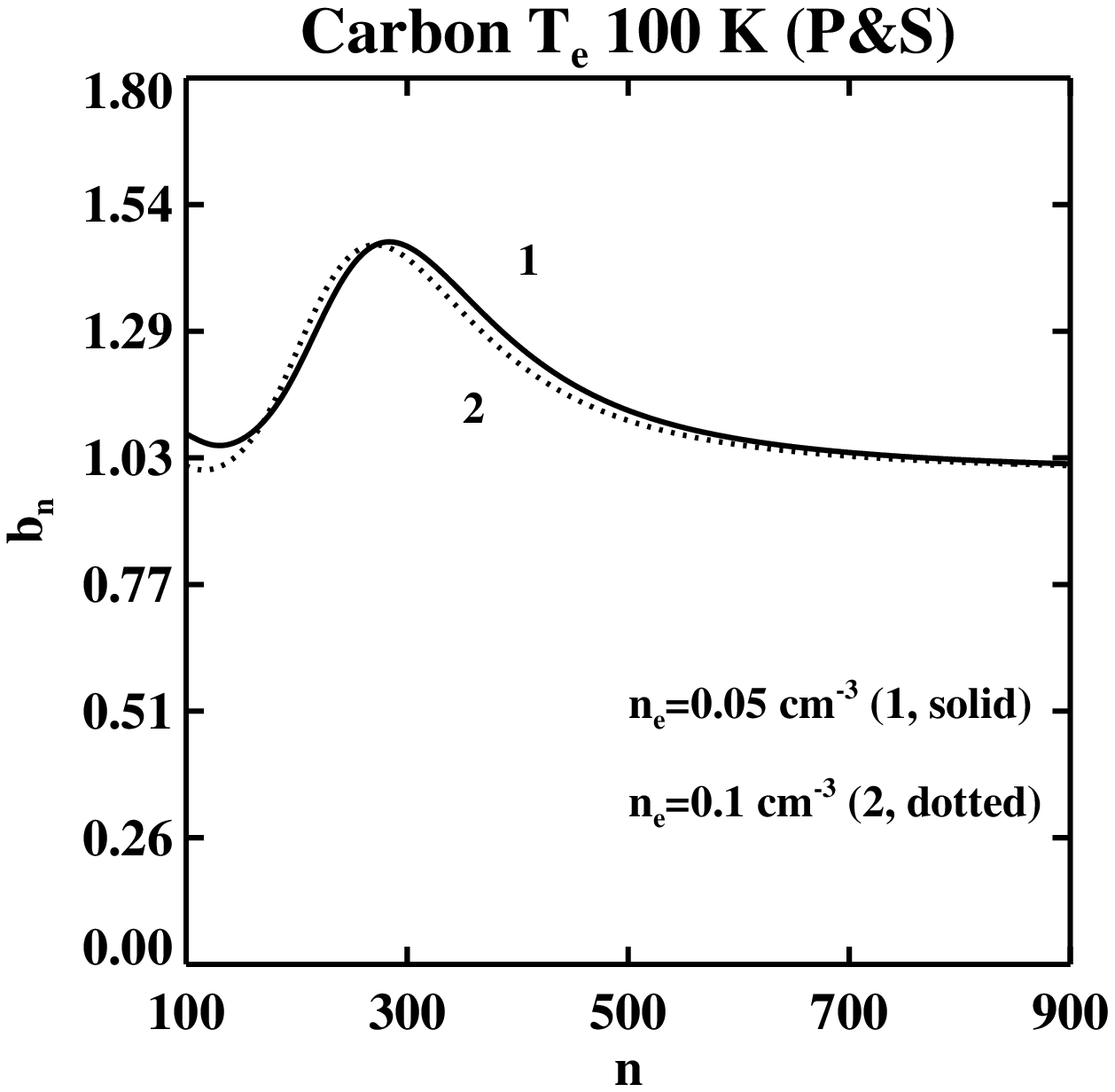}
\includegraphics[trim=0 0 110 0,clip=true,width=0.33\columnwidth]{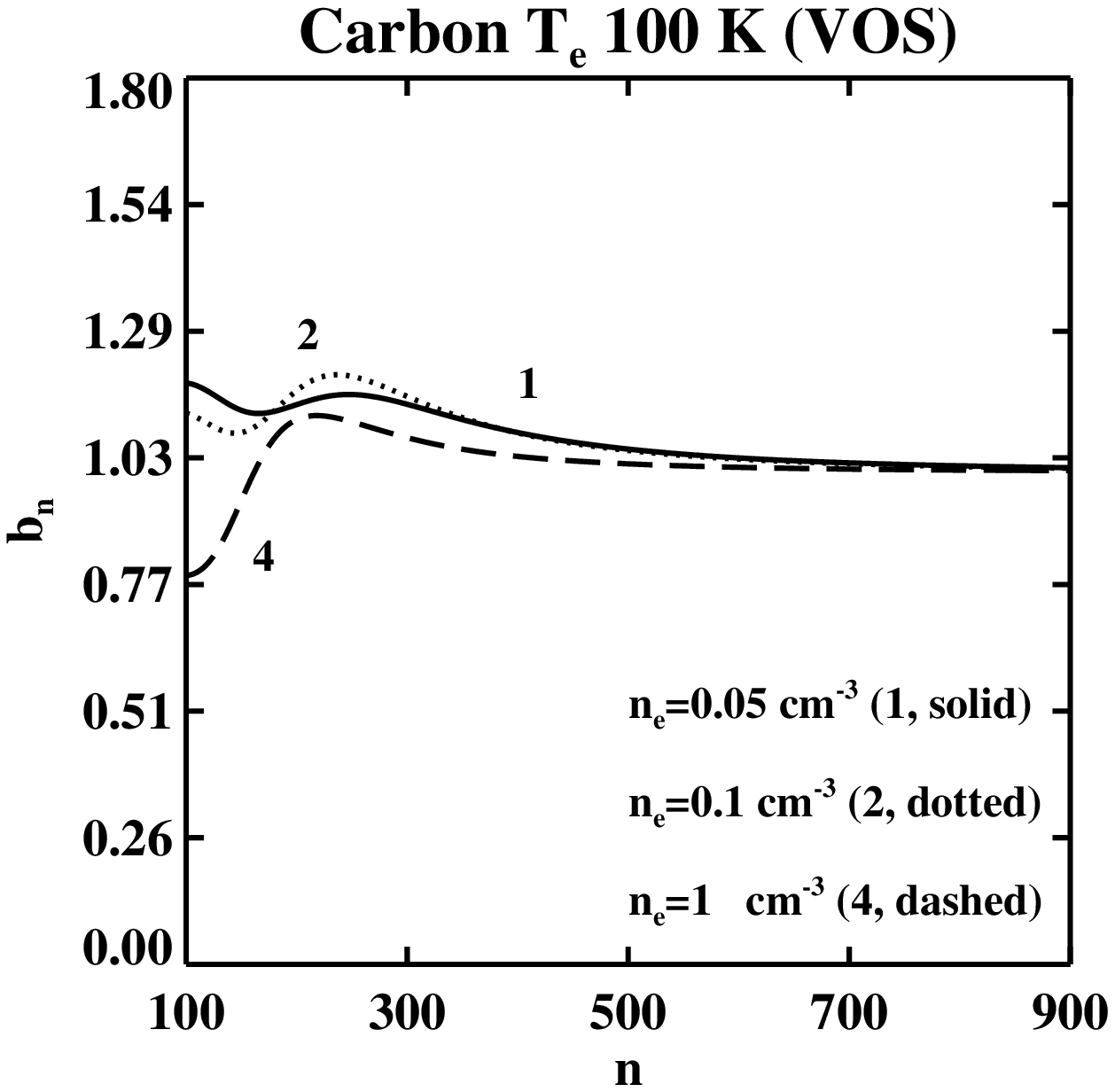}
\includegraphics[scale=1,bb=14 14 139 121,width=0.323\columnwidth,angle=-0.5, trim=0 -0.12cm 0 0]{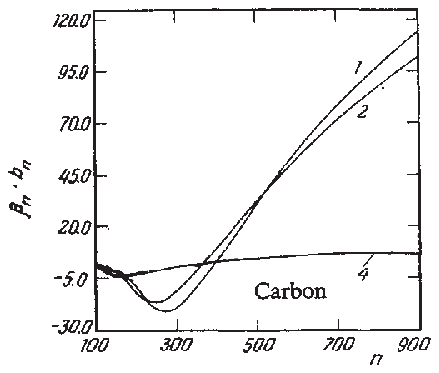}
\includegraphics[trim=0 0.1cm 110 0,clip=true,width=0.33\columnwidth]{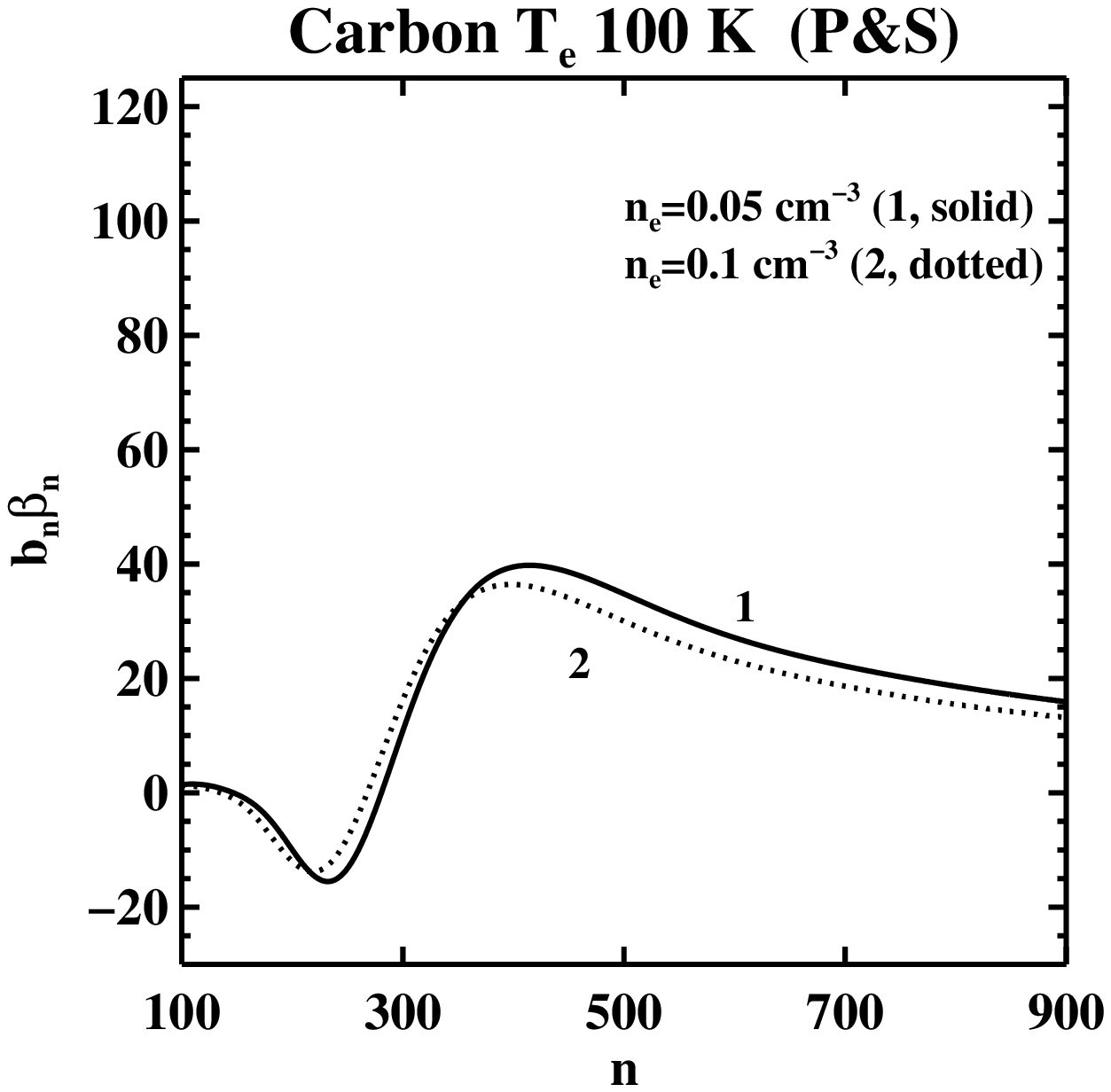}
\includegraphics[trim=0 0 110 0,clip=true,width=0.33\columnwidth]{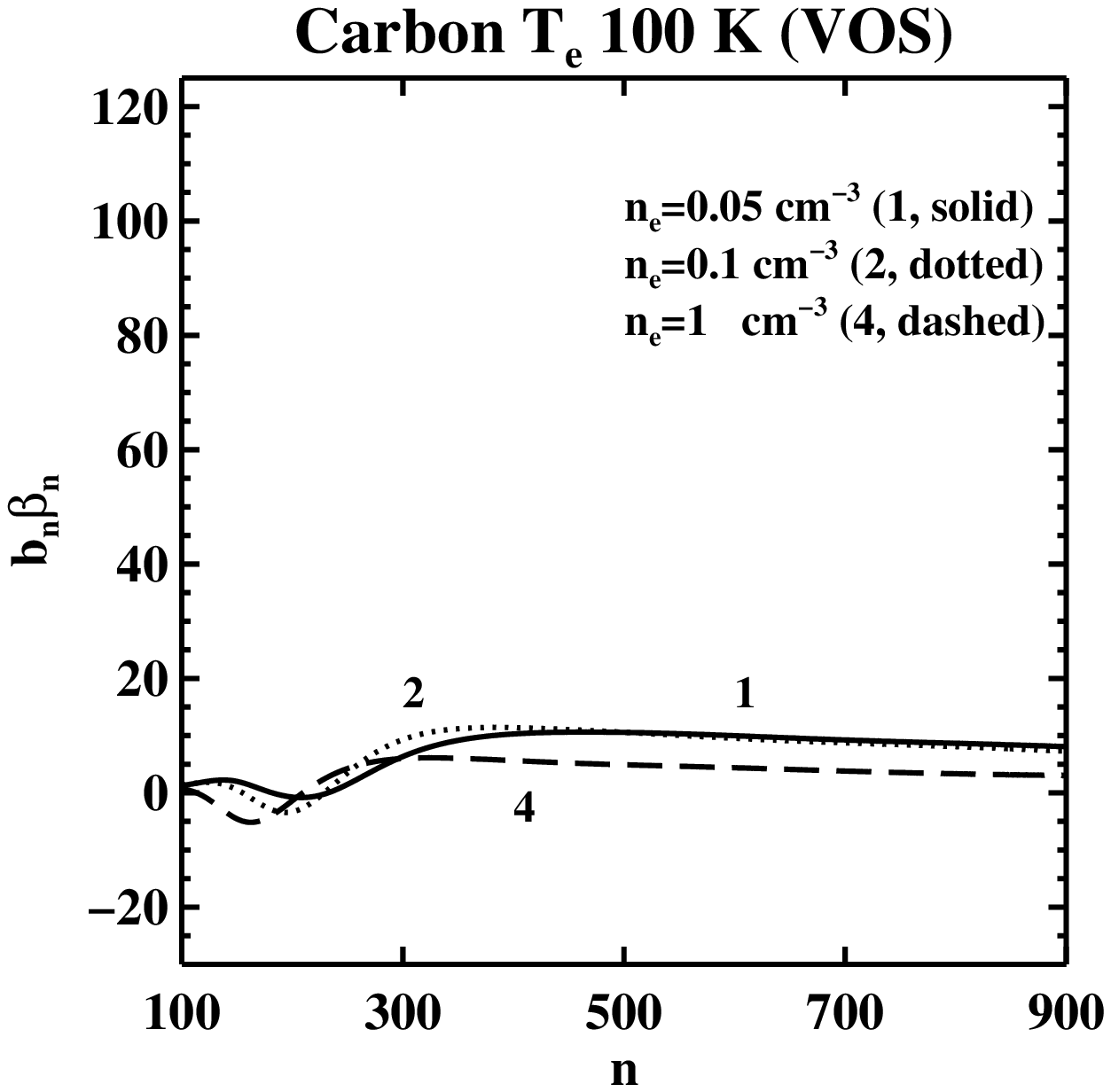}
\begin{picture}(10,10)
\put( 36,270){a)}
\put(186,270){b)}
\put(346,270){c)}
\put( 36,125){d)}
\put(186,125){e)}
\put(346,125){f)}
\end{picture}
\caption{Comparison between the CRRL departure coefficients from \citet{ponomarev1992}
(Panels a) and d); reproduced with permission from Ponomarev V.O. \& Sorochenko R. L., 1992, Soviet Astronomy Letters, 18, 215. Copyright 1992, AIP Publishing LLC.), this work using $l$-changing collision rates from \citet{pengellyandseaton1964} (Panels b) and e)) and those from \citet{vrinceanu2012} (Panels c) and f)) at $T_e$=100~K. Lines marked as 1, 2, 4 correspond to electron densities $n_e=$0.05, 0.1 and 1.0~cm$^{-3}$ (solid, dotted and dashed lines) respectively. The top panels show $b_n$ vs. $n$ and the bottom panels show the product $b_n\beta_n$ vs. $n$.\label{fig_compps2}}
\end{figure}

Now we compare departure coefficients obtained here with the results of \citet{ponomarev1992} and the effect of including $l$-changing collisions on the departure coefficients, see Figures~\ref{fig_compps} and \ref{fig_compps2}. We will focus the discussion on the $b_n$ values from \citet{ponomarev1992} as the \citet{walmsley1982} values are similar. 

While the results presented here are remarkably different from those of \citet{walmsley1982} and \citet{ponomarev1992}, some trends are similar.
We will first discuss the differences. Our results in Figures~\ref{fig_compps} and \ref{fig_compps2} show a pronounced 'bump' for low $n$ in the range 50 to 150. This bump is similar to what
we see for the hydrogenic approximation but enhanced by dielectronic recombination (c.f. Figures 3.6 and 8; Section 3.1.2). As discussed in Section 3.1.1 this bump
arises at these intermediate $n$ levels because collisions compete with spontaneous decay, effectively 'storing' electrons in high $l$ sublevels for
which radiative decay is less important. This means that the inclusion of $l$-changing collisions leads to significantly larger $b_{n}$ values
for $n$ in the range 50 to 150 as compared to \citet{ponomarev1992}. Regardless of the $l$-changing collision rates used, at higher $n$ we note that our $b_n$ values with increasing $n$ asymptotically
approach unity much faster than \citet{ponomarev1992}. This is especially true for lower electron densities ($n_e<$1.0~cm$^{-3}$) and
a direct consequence of using the $nl$-method to compute the departure coefficients.

Although the detailed behavior of our $b_{n}$ values differs strongly from \citet{ponomarev1992} there are also similarities in the general trends that we
observe as a function of electron density and temperature. In particular, the very low and very high $n$ asymptotic behavior of the $b_{n}$ values is similar to \citet{ponomarev1992}
in that the highest electron densities for a given electron temperature have the lowest $b_{n}$ values at low $n$ and approach equilibrium ($b_n$=1)
the fastest with increasing $n$. For higher electron densities and lower electron temperatures, our results become increasingly similar to the hydrogenic
case and agree with \citet{ponomarev1992}. This is expected as, as discussed in Section 3.1 at high densities the $b_n$ values approach equilibrium.

In terms of $b_n\beta_n$ our results show, as expected, good agreement with the hydrogenic case and \citet{ponomarev1992} in the high density and low
temperature limit. However, for the lower densities and higher temperatures shown in Figures~\ref{fig_compps} and \ref{fig_compps2} our models predict $b_n\beta_n$ values that are
lower by up to about an order of magnitude as compared to \citet{ponomarev1992}. This is particularly striking for the $T_e$=100~K and $n_e$=0.05~cm$^{-3}$
model shown in Figure~\ref{fig_compps2} where we find that both the maximum negative $b_n\beta_n$ value and maximum positive $b_n\beta_n$ value are more than
an order of magnitude lower than the corresponding \citet{ponomarev1992} values.

Since the integrated optical depth is directly proportional to the value of $b_n\beta_n$ (e.g. \citealt{salgado2016,walmsley1982,shaver1975}) we can interpret
$b_n\beta_n$ as a stimulation factor. This means that, for a given set of physical conditions, our models predict much lower maximum integrated optical
depths for Carbon as compared to earlier investigations (e.g. \citealt{walmsley1982,ponomarev1992}). This is true for both emission (negative $b_n\beta_n$) and
absorption (positive $b_n\beta_n$). In particular, our models predict that equilibrium will be reached at much lower $n$ (typically around $n=600$) and thus that
the integrated optical at high $n$ (low frequencies) will show a rather flat behavior for $n>$600 whereas the previous models by \citet{walmsley1982} and \citet{ponomarev1992}
predict a strong increase with increasing $n$.

\begin{figure}[!ht]
\begin{picture}(100,200)
\put(0,0){\includegraphics[width=0.5\columnwidth]{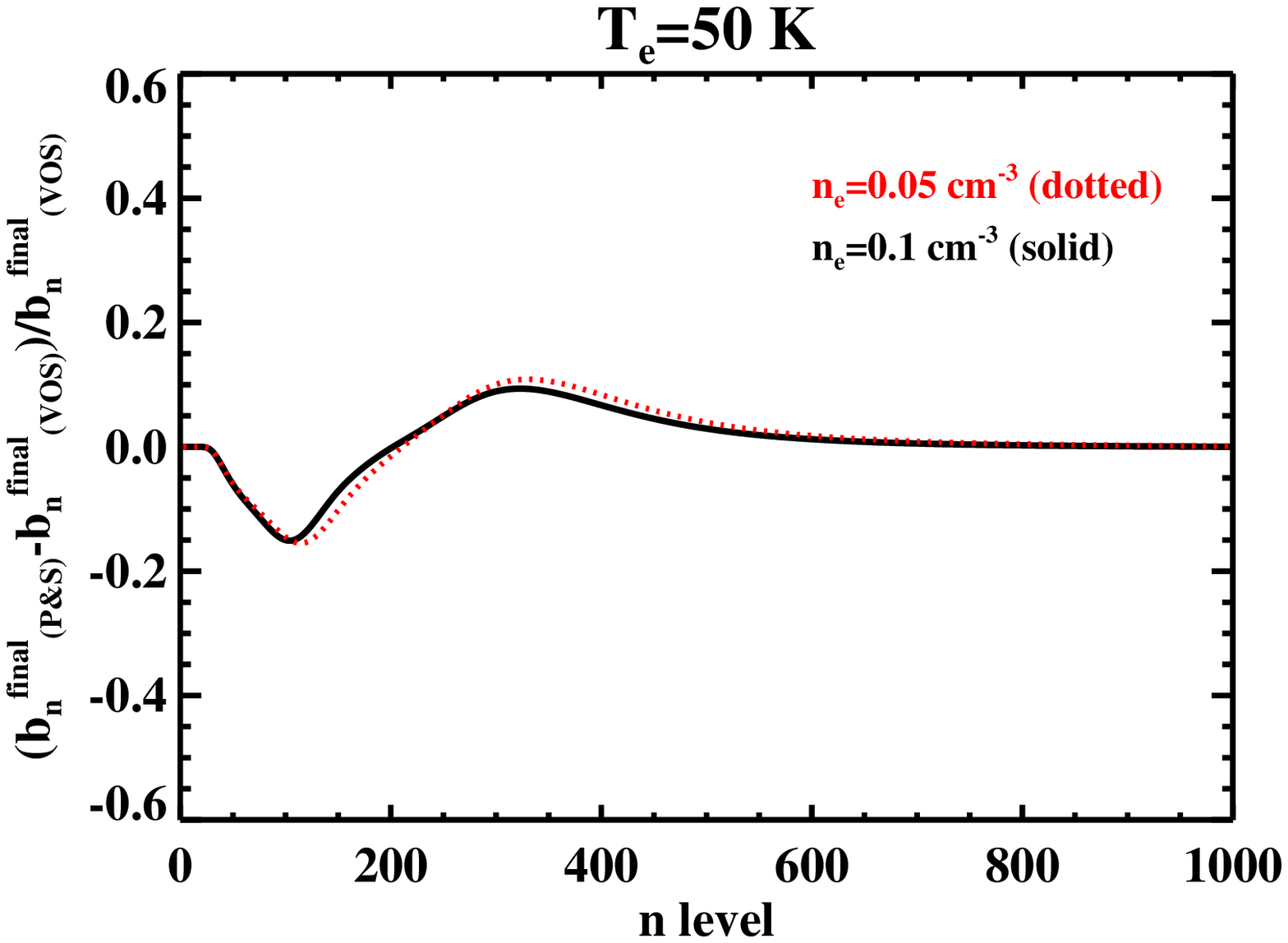}
\includegraphics[width=0.5\columnwidth]{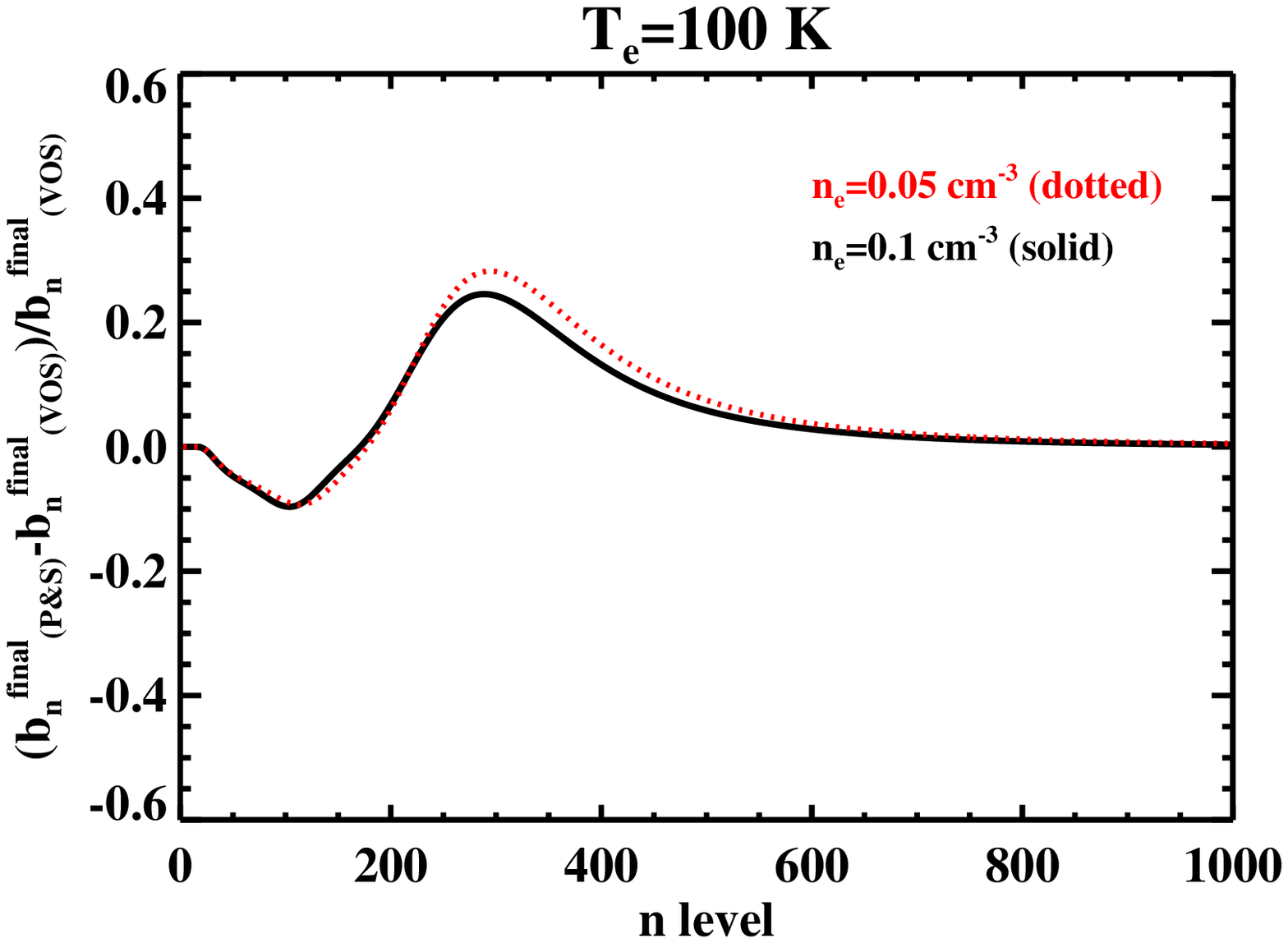}}
\put( 55,135){a)}
\put(295,135){b)}
\end{picture}
\caption{Comparison between the CRRL departure coefficients obtained using $l$-changing collisions rates from \citet{vrinceanu2012} and \citet{pengellyandseaton1964}
at 50 K and 100 K. The largest differences are $\sim30\%$ at levels $\sim300$.\label{fig_comppengelly}}
\end{figure}

We find that although our $b_n$ values asymptotically approach equilibrium at high $n$ that this value is not yet reached at $n=1000$. Therefore, the
$b_n\beta_n$ values we find are nearly, but not yet completely, constant in the range $n=$600-1000 and as such the dependence of integrated optical depth
on $b_n\beta_n$ remains important at high $n$. Finally, we note that for sufficiently high electron temperatures and low electron densities our models predict
the existence of a region at intermediate $n$ ($n=$100-200) where the $b_n\beta_n$ values can become positive. This behavior is a direct consequence of the
inclusion of $l$-changing collisions in our models. A more detailed comparison of the departure coefficients obtained using the $l$-changing collision from \citet{pengellyandseaton1964} and those using the rates from \citet{vrinceanu2012} (Figure \ref{fig_comppengelly}) reveals differences of less than 30\% for the conditions of interest for CRRL studies.

Apart from the $l$-changing collisions there are other potentially important differences between our models and those published by \citet{ponomarev1992}. \citet{ponomarev1992}
do not provide the explicit values of the dielectronic recombination rates that they use. However, they refer back to \citet{walmsley1982} for these rates and as we use the same formalism, we do not think that the dielectronic recombination rates are at the heart of the discrepancy. may have influenced their results.
In addition, we note that we use somewhat different collision rates in our simulations. However, as illustrated in Figure 13 and 14, the exact collision rates have only limited influence on the $b_n$ values. Rather, we suspect that the approximate way the statistical equilibrium equations are solved by \citet{ponomarev1992} may have influenced their results and that including $l$-changing collisions properly rather than adopting a statistical populations as did \citet{ponomarev1992} is key.

\begin{figure}[!ht]
\includegraphics[width=1\columnwidth]{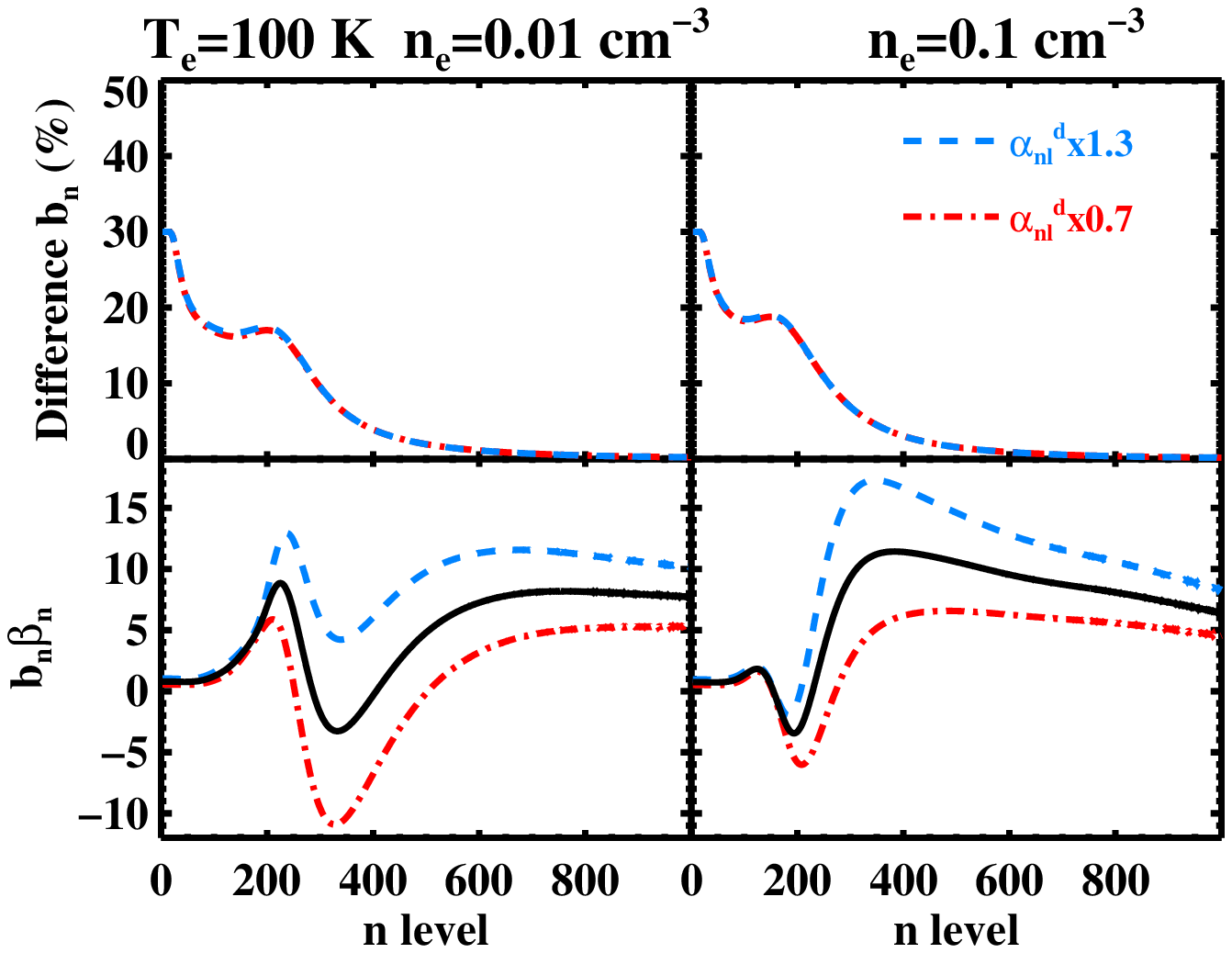}
\caption{Upper panels: the difference (in percentage) of the departure coefficients after multiplying the dielectronic recombination rate by a factor of
1.3 (0.7) in blue (red) for a temperature of 100 K and densities of 0.01~$\mathrm{cm^{-3}}$ (left) and 0.1~$\mathrm{cm^{-3}}$ (right). Lower panels:
$b_n \beta_n$ for the same physical conditions as in the upper panels. At $n$ levels lower than $\sim200$ the $b_n \beta_n$ values derived by using 
the modified dielectronic recombination rates (blue and red) are similar to those without modification (black). At higher levels the overall trends
are similar but they can differ by factors of a few.\label{fig_effectofdielec}}
\end{figure}

A further assessment of the effect of any uncertainty in the adopted dielectronic recombination rates on the final departure coefficients can be performed by
arbitrarily multiplying the dielectronic recombination rate by a factor. We note that a dielectronic recombination rate a factor of 30\% higher (lower)
increases (decreases) the departure coefficients at low levels ($n < 100$) by 30\%. At the higher levels of interest for the study of CRRLs, ($n > 250$)
a factor of 30\% on the dielectronic recombination rates changes the values of the departure coefficients by less than 10\% (Figure \ref{fig_effectofdielec}, upper panels).
As expected, the values for $b_n\beta_n$ are affected more by the change on the dielectronic recombination rate and can be altered by a factors of a
few (Figure \ref{fig_effectofdielec}, lower panels). It is clear that quantitative interpretation of carbon radio recombination lines would be served by more accurate
dielectronic recombination rates that include the fine structure levels.

\section{Summary and Conclusions}\label{section_conclusions}

We have solved the level population equation for hydrogenic atoms using novel rates involved in the process.
The level population equation is solved in two approximations: the $n$ and the $nl$ method. The departure coefficients obtained
using the $n$ method are similar to values from the literature (e.g. \citealt{brocklehurst1970} and \citealt{shaver1975}).
Our results using the $nl$ method reproduce those from \citet{hummer1987} well, once allowance is made for updates in the collisional rates.

By including the dielectronic recombination process together with the $nl$ method we are able to model
the level population of carbon in terms of the departure coefficients. Our results are qualitatively similar to those of \citet{watson1980,walmsley1982}.
However, the values obtained here differ considerably from those from the literature. The differences can be
understood in terms of the use of improved collision rates and the improved numerical approach using the $nl$ method.
We confirm that dielectronic recombination can indeed produce an increase on the values of the departure coefficients
at high $n$ levels compared to the hydrogenic values.

In anticipation of low frequency radio recombination line surveys of the diffuse interstellar medium now being undertaken
by LOFAR, we have expanded the range of applicability of the formulation to the conditions of the cold neutral medium. For this environment,
external radiation fields also become important at intermediate principal quantum levels while at high levels the influence of radiation
fields on the level population is less important In an accompanying paper \citep{salgado2016}, we discuss the expected line strength
for low frequency carbon radio recombination lines and the influence of an external radiation field. Throughout this work we have used a zero
radiation field. In this companion paper we compare our results to existing observations of CRRLs towards Cas~A and regions in the
inner galaxy. We also describe the analysis techniques and diagnostic diagrams that can be used to analyze the forthcoming LOFAR CRRL survey.
The departure coefficients obtained here will be used to analyze the LOFAR observations of Cas~A in a future article \citep{oonk2015b}.

\clearpage

\appendix

\section{List of Symbols}\label{appendix_listofsymbols}

\begin{center}
\begin{longtable}{lp{100mm}}
\caption{List of Symbols\label{table_listofsymbols}}\\ 
\multicolumn{2}{l}{}\\
\hline
\hline
Symbol & Descritpion \\
\hline
\endfirsthead
\multicolumn{2}{l}
{\tablename\ \thetable\ -- \textit{Continued from previous page}} \\
\hline
\hline
Symbol &  \\
\hline
\endhead
\hline \multicolumn{2}{r}{\textit{Continued on next page}} \\
\endfoot
\endlastfoot
 $A_{3/2,1/2}$& Spontaneous transition rate of the carbon fine structure line ${^2}P_{3/2}$-${^2}P_{1/2}$\\
 $A^a_{nl}$ & Autoionization rate\\
 $A_{n'n}$ & Einstein coefficient for spontaneous transition between $n'$ and $n$\\
 $A_{n'l'nl}$ & Einstein coefficient for spontaneous transition between $n'l'$ state to $nl$ state\\
 $a_0$ & Bohr radius\\
 $a_{nl}$ & Photoionization cross section\\
 $B_{nn'}$ & Einstein coefficient for stimulated transition from level $n'$ to $n$\\
 $b_n$& Departure coefficient for level n\\
 $b_n^{1/2}$ & Departure coefficient for atoms recombining from the $1/2$ ion core for level $n$\\
 $b_n^{3/2}$ & Departure coefficient for atoms recombining from the $3/2$ ion core for level $n$\\
 $b_n^{final}$ & Departure coefficient for atoms recombining from both ion cores\\
 $\mathrm{C}n\alpha$& Carbon recombination line for $\alpha$ transition\\
 $C_{n'n}$& Rates for energy changing collisions between level $n'$ and $n$\\
 $C(n,l)$& Coefficient for recursion relations used to obtain the radial matrices values \\
 $c$& Speed of light\\ 
 $EM_{\mathrm{C+}}$& Emission measure of carbon ions\\
 $g_{3/2}$ & Statistical weight for the fine structure level ${^2}P_{3/2}$\\
 $g_{1/2}$ & Statistical weight for the fine structure level ${^2}P_{1/2}$\\
 $h$& Planck constant\\
 $I_0(\nu)$& Intensity of the background continuum\\
 $I_\nu^{line}$& Intensity of the line\\
 $I_\nu^{cont}$& Intensity of the continuum\\
 $I_{158}$& Intensity of the fine structure line of carbon at 158~$\mathrm{\mu m}$\\
 $j_\nu$ & line emission coefficient\\
 $k_\nu$ & line absorption coefficient\\
 $k$& Boltzmann constant\\
 $L$& Pathlength of cloud\\
 $l$& Angular momentum quantum number\\
 $N_{cr}$& Critical density for collisions on a two level atom\\
 $N_n$& Density of atoms in level $n$\\
 $N_{nl}$& Density of atoms in level $n$ and sublevel $l$\\
 $N_e$& Electron density\\
 $N_H$& Hydrogen density\\
 $N_{ion}$ & Density of the parent ions \\
 $N_{3/2}^+$& Level population of carbon ions in the ${^2}P_{3/2}$ core\\
 $N_{1/2}^+$& Level population of carbon ions in the ${^2}P_{1/2}$ core\\
 $n$& Lower principal quantum number\\
 $n'$& Upper principal quantum number\\
 $n_{max}$ & Maximum level considered in our simulations\\
 $n_{crit}$ & Critical level considered in our simulations for the $nl$-method\\
 $n_t$& Level where observed lines transition from emission to absorption\\
 $\mathscr{R}(n,l)$ & Normalized radial wave function for level $n$, $l$\\
 $R$ & Ratio between the fine structure (${^2}P_{3/2}$-${^2}P_{1/2}$) level population and the fine structure level population in LTE\\
 $R(l',l)$ & Integral of the radial matrix elements\\
 $Ry$ & Rydberg constant\\
 $T_0$& Temperature of power law background spectrum at frequency $\nu_0$\\
 $T_e$& Electron temperature\\
 $Z$ & \\
 $\alpha_n$& Radiative recombination coefficient to a level $n$\\
 $\alpha_{nl}$& Radiative recombination coefficient to a level $n$ and sublevel $l$\\
 $\alpha^d_{nl}$& Dielectronic recombination rate\\
 $\beta_{n n'}$& Correction factor for stimulated emission\\
 $\gamma_e$& De-excitation rate for carbon ions in the ${^2}P_{3/2}$ core due to collisions with electrons\\ 
 $\gamma_H$& De-excitation rate for carbon ions in the ${^2}P_{3/2}$ core due to collisions with hydrogen atoms\\
 $\Delta E$& Energy difference between two levels\\
 $\Delta n$& $n'-n$, difference between the upper an lower principal quantum number\\
 $\eta$& Correction factor to the Planck function due to non-LTE level population\\
 $\mu$& Reduced mass\\
 $\nu$& Frequency of a transition\\
 $\nu_0$& Reference frequency for the power law background spectrum\\
 $\phi(\nu)$& Line profile\\
 $\omega_{nl}$& Statistical weight of level $nl$\\
 $\omega_i$& Statistical weight of parent ion\\
 $\chi_n$ & Ionization potential of a level $n$, divided by $k T_e$\\
\hline
\end{longtable} 
\end{center}

\section{Level population}
The strength (or depth) of an emission (absorption) line depends on the level population of atoms.
The line emission and absorption coefficients are given by (e.g. \citealt{shaver1975,gordon2009}):
\begin{eqnarray}
j_\nu &=& \frac{h \nu}{4 \pi} A_{n'n} N_{n'} \phi(\nu),\\
k_\nu &=& \frac{h \nu}{4 \pi} \left( N_{n} B_{nn'}-N_{n'} B_{n'n}  \right) \phi(\nu),
\end{eqnarray}
\noindent where $h$ is the Planck constant, $N_{n'}$ is the level population of a given upper level ($n'$) and $N_{n}$ is the level
population of the lower level ($n$); $\phi(\nu)$ is the line profile, $\nu$ is the frequency of the transition and $A_{n'n}$,
$B_{n'n}(B_{nn'})$ are the Einstein coefficients for spontaneous and stimulated emission (absorption), respectively.
Following \citet{hummer1987}, we present the results of our modeling in terms of the departure coefficients ($b_n$) and the correction factor
for stimulated emission/absorption ($\beta_n$):
\begin{eqnarray}\label{eqn_bnapp}
b_n=\frac{N_n}{N_n(LTE)}.
\end{eqnarray}
\begin{eqnarray}\label{eq_betaapp}
\beta_{n,n'}=\frac{1-\left(b_{n'}/b_n\right) \exp(-h \nu/k T_e)}{1-\exp(-h \nu/k T_e)},
\end{eqnarray}
\noindent unless otherwise stated the $\beta_n$~presented here correspond to $\beta_{n+1,n}$, i.e. $\alpha$~transitions.
When a cloud is located in front of a strong background source the integrated line to continuum ratio is proportional to $b_n \beta_n$ \citep{shaver1975,payne1994}.
We expand on the radiative transfer problem in Paper II.

\subsection{Hydrogenic atoms}\label{section_levelpopH}

Under thermodynamic equilibrium conditions, level populations are given by the Saha-Boltzmann equation (e.g. \citealt{brocklehurst1972,gordon2009}):
\begin{eqnarray}
N_{nl}(LTE)&=&N_e N_{ion}\left(\frac{h^2}{2 \pi m_e k T_e}\right)^{3/2} \frac{\omega_{nl}}{2\omega_i} e^{\chi_n}, \chi_n=\frac{hc Z^2 Ry}{n^2kT_e},
\end{eqnarray}
\noindent where $N_e$~is the electron density in the nebula, $N_{ion}$~is the ion density, $m_e$~is the electron mass, $k$ is the Boltzmann constant,
$Ry$ is the Rydberg constant $\omega_{nl}$~is the statistical weight of the level $n$ and angular quantum momentum level $l$~[$\omega_{nl}=2(2l+1)$,
for hydrogen], $\omega_i$ is the statistical weight of the parent ion. The factor $\left({h^2}/{2 \pi m_e k T_e}\right)^{0.5}$~is
the thermal de Broglie wavelength, $\Lambda$, of the free electron $\left[\Lambda(T_e)^3\approx4.14133\times10^{-16}~T_e^{-1.5}~\mathrm{cm^3}\right]$.
In general, lines are formed under non-LTE conditions and, in order to properly model the line behavior, the level population equation must
be solved. We follow the methods described in \citet{brocklehurst1971} and improved upon by \citet{hummer1987} as described in Section \ref{section_method}.
Here, we give a detailed derivation of the theory and methods. First, we solve the level population equation assuming statistical population of the angular
momentum $l$-levels, i.e.:
\begin{eqnarray}\label{eqn_neqnl}
N_n=\sum_{l=0}^{n-1}\frac{(2l+1)}{n^2} N_{nl},
\end{eqnarray}
\noindent for all $n$~levels. This assumption greatly simplifies the calculations but is only valid when $l$~changing transitions are faster than other processes,
and, in general, this is not the case for low $n$~levels. The level population equation under this assumption is (e.g. \citealt{shaver1975, gordon2009}):
\begin{eqnarray}\label{eqn_levelpop1}
N_n\left[\sum\limits_{n' < n}A_{nn'}+\sum\limits_{n' \neq n}{(B_{nn'}I_\nu + C_{nn'})}+ C_{ni}\right] &=& \sum\limits_{n'> n}{N_{n'} A_{n'n}}+\sum\limits_{n' \neq n}{N_{n'}(B_{n'n}I_\nu +C_{n'n})} \nonumber\\
&&+N_e N_{ion}(\alpha_n+ C_{in}).
\end{eqnarray}
\noindent The right- and left-hand side of Equation~\ref{eqn_levelpop1} describe how level $n$~is populated and depopulated, respectively.
We take into account spontaneous transitions from level $n$~to lower levels ($A_{nn'}$), stimulated emission and absorption ($B_{nn'}I_\nu$, $B_{n'n}I_\nu$), collisional transitions ($C_{nn'}$),
radiative recombination ($\alpha_n$), collisional ionization ($C_{in}$) and 3-body recombination ($C_{ni}$).
Equation~\ref{eqn_levelpop1} can be written in terms of the departure coefficients ($b_n$):
\begin{eqnarray}\label{eqn_levelpop2}
b_n\left[\sum\limits_{n' < n}A_{nn'}+\sum\limits_{n' \neq n}{(B_{nn'}I_\nu + C_{nn'})}+ C_{ni}\right] &=& \sum\limits_{n' > n}{b_{n'} \frac{\omega_{n'}}{\omega_n} e^{\Delta \chi_{n'n}} A_{n'n}}\nonumber\\
 +\sum\limits_{n' \neq n}{b_{n'} \frac{\omega_{n'}}{\omega_n} e^{\Delta \chi_{n'n}}(B_{n'n}I_\nu +C_{n'n})}+\frac{N_e N_{ion}}{N_n(LTE)}(\alpha_n+ C_{in}).
\end{eqnarray}
\noindent The previous equation can be written as a matrix equation of the form {\bf R}$\times${\bf b}={\bf S} by choosing the appropriate elements
to form the matrices {\bf R} and {\bf S} (e.g. \citealt{shaver1975}):
\begin{eqnarray}
R_{n'n}&=& -\frac{\omega_{n'}}{\omega_n} e^{\Delta \chi_{n'n}} (A_{n'n}+B_{n'n}I_\nu +C_{n'n}), (n'>n) \\
R_{nn}&=& \sum\limits_{n' < n}A_{nn'}+\sum\limits_{n' \neq n}{(B_{nn'}I_\nu + C_{nn'})}+ C{_n}{_i} \\
R_{n'n}&=& -\frac{\omega_{n'}}{\omega_n} e^{\Delta \chi_{n'n}}(B_{n'n}I_\nu +C_{n'n}), (n'<n) \\
S_{n}&=& \frac{N_e N_{ion}}{N_n(LTE)}(\alpha_n+ C{_i}{_n}).
\end{eqnarray}
It is easy to solve for the $b_n$~values by using standard matrix inversion techniques. We will refer to this approach of solving the level population
equation as the $n$-method.

At low $n$~levels, the quantum angular momentum distribution must be obtained, since the assumption that the angular momentum levels are in statistical equilibrium
is no longer valid. Moreover, as described in \citet{watson1980, walmsley1982}, dielectronic recombination is an important process for carbon ions at low
temperatures and densities. Since the dielectronic recombination process depends on the quantum angular momentum distribution, we need to include the $l$ sublevel
distribution for a given $n$~level.

The level population equation considering $l$-levels is:
\begin{eqnarray}\label{eqn_llevelpop}
b_{nl}\left[\sum\limits_{n' < n}\sum_{l'=l\pm1}A_{nln'l'}+\sum\limits_{n' \neq n}{(B_{nln'l'}I_\nu + C_{nln'l'})}+\sum_{l'=l\pm1}C_{nlnl'}+ C_{nl,i}\right] &=&\nonumber \\
 \sum\limits_{n' > n}\sum_{l'=l\pm1}{b_{n'l'} \frac{\omega_{n'l'}}{\omega_{nl}} e^{\Delta \chi_{n'n}} A_{n'l'nl}}+\sum\limits_{n' \neq n}\sum_{l'=l\pm1}{b_{n'l'} \frac{\omega_{n'l'}}{\omega_{nl}} e^{\Delta \chi_{n'n}}(B_{n'l'nl}I_\nu +C_{n'l'nl})} \nonumber\\ 
+\sum_{l'=l\pm1} b_{nl'}\left(\frac{\omega_{nl'}}{\omega_{nl}}  \right)C_{nl'nl}+\frac{N_e N_{ion}}{N_{nl}(LTE)}(\alpha_{nl}+ C_{i,nl}).
\end{eqnarray}
To solve for the $l$~level distribution at a given $n$~level we followed an iterative approach as described in \citet{brocklehurst1971,hummer1987}.
We will refer to this approach of solving the level population equation as the $nl$-method.

We start the computations by applying the $n$-method, i.e. assuming $b_{nl}=b_n$~for all $l$~levels, thus obtaining $b_{nl}^{(0)}$~values. For levels
above a given $n_{crit}$ value we expect the $l$-sublevels to be in statistical equilibrium. In this case, Equation~\ref{eqn_neqnl} is valid
and the $b_{nl}$~values are equal to those obtained by the $n$-method.
On the first iteration, we start solving Equation~\ref{eqn_llevelpop} at $n=n_{crit}$ and use the previously computed values ($b_{n'l'}^{(0)}$)~for levels $n'\neq n$.
Equation~\ref{eqn_llevelpop} is then a tri-diagonal matrix (only elements with $l'=l\pm1$, enter in the equation) and, by solving
the system of equations, we obtain $b_{nl}^{(1)}$~values. The operation is repeated for all $n$~levels down to $n=n_{min}$.
In all our simulations we assume $n_{min}=3$ since we are focused on studying carbon atoms whose ground level correspond
to $n=2$. We repeat the operation by using the $b_{nl}^{(1)}$~values instead of the $b_{nl}^{(0)}$~values. \citet{hummer1987} have proven that
considering collisions from (and to) all $n'$~levels guarantees a continuous distribution between both approaches at levels close to $n_{crit}$.
The final $b_n$~values are computed by taking the weighted sum of the $b_{nl}$ values:
\begin{eqnarray}\label{eqn_bn2app}
b_n=\sum_{l=0}^{n-1} \left(\frac{2l+1}{n^2}\right)b_{nl},
\end{eqnarray}
\noindent Details on the parameters used in this work are given in the text (Section~\ref{section_nummethod}).

\section{Radial Matrices and Einstein A coefficients}\label{app_einsa}
In general, the radiative decay depends on the angular momentum quantum number of the electron at the level $n$.
Transitions from level $nl\rightarrow n' l'$ are described by $A_{nln'l'}$~coefficients, in the dipole
approximation \citep{seaton1959a}:
\begin{eqnarray}
A_{nln'l'}&=& \frac{64 \pi^4 \nu^3}{3 h c^3}e^2 a_0^2 \frac{\mathrm{max}(l,l')}{2 l+1} \left|\int\limits_{0}\limits^\infty \mathscr{R}(n',l')r \mathscr{R}(n,l) \mathrm{d}r\right|^2,
\end{eqnarray}
\noindent where $a_0$~is the Bohr radius and $\mathscr{R}(n,l)$~is the normalized radial wave function solution to the Schr\"odinger equation of the Hydrogen atom
\citep{burgess1958,brocklehurst1971}.
The computation of the matrix elements is challenging (see \citealt{morabito2014a} for details) and we follow the recursion relations given by
\citet{storey1991} to calculate them up to $n=10000$. Defining:
\begin{eqnarray}
R(l',l) = \int\limits_{0}\limits^\infty \mathscr{R}(n',l')r \mathscr{R}(n,l) \mathrm{d}r,
\end{eqnarray}
\noindent where the first argument of $R(l',l)$~corresponds to the lower state. For a given $n'$~level, \citet{storey1991} give the following relations,
with the starting values:
\begin{eqnarray}
R(n',n'-1)&=&0,\\
R(n'-1,n')&=&\frac{1}{4}\left(4nn' \right)^{n'+2}\left[ \frac{(n+n')!}{(n-n'-1)!(2n'-1)!} \right]^{1/2} \frac{(n-n')^{n-n'-2}}{(n+n')^{n+n'+2}}.
\end{eqnarray}
The recursion relations are:
\begin{eqnarray}
2lC(n',l) R(l-1,l) = (2l+1)C(n,l+1)R(l,l+1)+C(n',l+1)R(l+1,l),
\end{eqnarray}
\noindent and:
\begin{eqnarray}
2lC(n,l) R(l,l-1) = C(n,l+1)R(l,l+1)+(2l+1)C(n',l+1)R(l+1,l),
\end{eqnarray}
\noindent with:
\begin{eqnarray}
C(n,l)=\frac{\sqrt{(n+l)(n-l)}}{nl}
\end{eqnarray}

\section{Radiative recombination cross-section}\label{app_radrec}
\citet{storey1991} give a formula for computing the photoionization cross-section:
\begin{equation}
a_{nl}(h \nu)= \left(\frac{4\pi a_0^2 \alpha}{3}\right) \frac{\left(1+n^2 \kappa^2\right)}{\mu^2 Z^2 n^2} \frac{\mathrm{max}(l,l')}{2l+1} \left|\int\limits_{0}\limits^\infty \mathscr{R}(n',l')r \mathscr{R}(\kappa,l)\mathrm{d}r\right|^2.
\end{equation}
\noindent To obtain the radial matrices elements, we use the same recursion formula as for the Einstein A coefficients with the substitution:
$n = i/\kappa$, with $i$~the imaginary number. The $C(n,l)$ coefficients are:
\begin{eqnarray}
C(n,l)=\frac{\sqrt{(1+l^2\kappa^2)}}{l},
\end{eqnarray}
\noindent and the initial values are:
\begin{eqnarray}
R(n',n'-1)&=&0, \nonumber\\
R(n'-1,n')_{\kappa=0}&=&\frac{1}{4} \left[ \frac{\pi}{2\left(2n'-1 \right)!}\right]^{1/2} (4n')^{n'+2}e^{-2n'},\nonumber\\
R(n'-1,n')_{\kappa \neq 0}&=& \left[\frac{\prod_{s=1}^{n'} (1+s^2\kappa^2)}{1-\mathrm{exp}(-2\pi/\kappa)} \right]^{1/2} \frac{ \mathrm{exp}[2n'-(2/\kappa) \mathrm{arctan}(n'\kappa)]}{\left(1+n'^2\kappa^2\right)^{n'+2}}  R(n'-1,n')_{\kappa=0}\nonumber.
\end{eqnarray}

We are interested in computing the recombination cross-section for an electron with energy $E$~recombining to a level $nl$. From
Milne relation we obtain (e.g. \citealt{rybicki1986}):
\begin{eqnarray}
\sigma(E,nl)= \left(\frac{16\pi a_0^2}{3\sqrt{2}}\right) \sqrt{\left(\frac{hc Ry}{E}\right)} \sqrt{\left(\frac{m_e c^2}{E} \right)} \left( \frac{E+ h\nu_n} {m_e c^2}\right)^3 \sum_{l'} \mathrm{max}(l,l') \left|\int\limits_{0}\limits^\infty \mathscr{R}(\kappa,l')r\mathscr{R}(n,l)\mathrm{d}r\right|^2,
\end{eqnarray}
\noindent expressed in terms of the radial matrices. Here, $h \nu_n$~is the ionization energy of the level $n$.
The final rate is obtained by integrating the cross-section over a Maxwellian velocity distribution:
\begin{eqnarray}
\alpha_{nl}= \frac{8}{\sqrt{\pi m_e}} (k T_e)^{-3/2} \int\limits_0\limits^\infty \sigma(E,nl) e^{-E/kT_e} \mathrm{d}E.
\end{eqnarray}
\noindent We consider $x=E/kT_e$~and $I(x)$ is the function in the integral.
To integrate the cross-section, we followed an approach similar to \citet{burgess1965}. We divide the integral in 30 segments starting at $x_0=kT \times 10^{-10}$,
and ending at $x_f=20 \times kT$. Each segment is integrated by using a 6-point Gauss-Legendre quadrature scheme. This approach provides
the value of the integral close to $kT$, therefore two correction factors must be applied: for the small values of $x$~we note that the integrand
is almost constant and the value of the integral is then $I(x_0)x_0^2/2$; for large values of $x$ we use a 6-point Gauss-Legendre quadrature
starting at $x_0=20 \times kT $~and ending at $x_f=30\times kT$. As mentioned in Section~\ref{section_results} we compare the sum over $l$,
of our radiative recombination rates with the formula of \citet{seaton1959b}:
\begin{eqnarray}
\alpha{_n}=2.06\times10^{-11}\left( \frac{Z}{n T_e^{0.5}}\right) \chi_n S_n(\lambda)~\mathrm{cm^3 s^{-1}},
\end{eqnarray}
\noindent with $\lambda=n^2 \chi_n$, and
\begin{eqnarray}
S_n(\lambda) = \int \limits_{0} \limits^\infty \frac{g_{II}(n,\epsilon) e^{-x_n u}}{1+u} du, u=n^2 \epsilon.
\end{eqnarray}
Values for the $\chi_n S_n(\lambda)$~are given by \citet{seaton1959b} in two approximations for large and small argument, and
tabulated values are also given for values in between the approximations. A first order expansion of the Gaunt factor \citep{allen1973}
provides an accurate formula for the recombination coefficient:
\begin{eqnarray}\label{eqrecomb}
\alpha_n=3.262\times10^{-6}\left( \frac{Z^4}{n^3 T_e^{1.5}}\right) e^{\chi_n} E_1\left(\chi_n \right)~\mathrm{cm^3 s^{-1}}.
\end{eqnarray}

\section{Energy changing collision rates}\label{app_cnnp}
\citet{vriens1980} obtained the following semi-empirical formula for excitation by electrons. The formula is given by:
\begin{eqnarray}
C_{nn'}=1.6\times10^{-7}\frac{\sqrt{kT_e}}{kT_e+\Gamma_{nn'}}\exp(-\epsilon_{nn'}) \left[ A_{nn'} \ln(0.3 \frac{kT_e}{h c Ry}+\Delta_{nn'})+B_{nn'}\right],
\end{eqnarray}
\noindent with the coefficients defined as:
\begin{align*}
&s=|n-n'|, \\
&A_{nn'}=2 \frac{h c Ry}{E_{nn'}} f_{nn'}, \\
&B_{nn'}=4 \frac{(h c Ry)^2}{n^3}\left( \frac{1}{E_{nn'}^2}+\frac{4}{3}\frac{E_{ni}}{E_{nn'}^3}+b_p \frac{E_{ni}^2}{E_{nn'^4}}  \right), \\
&b_p=1.4\frac{\mathrm{ln}(n)}{n}-\frac{0.7}{n}-\frac{0.51}{n^2}+\frac{1.16}{n^3}-\frac{0.55}{n^4}, \\
&\Delta_{nn'}=\exp(-\frac{B_{nn'}}{A_{n,n'}})+0.06\frac{s^2}{n n'^2}, \\
\end{align*}
\begin{align*}
\Gamma_{nn'}= h c Ry \ln \left(1+\frac{n'^3 kT_e}{h c Ry} \right)\left[ 3+11\left( \frac{s}{p}\right)^2 \right] \left( 6+1.6 n s +\frac{0.3}{s^2}+0.8\sqrt{\frac{n^3}{s}}\left|s-0.6\right| \right)^{-1}.\\
\end{align*}

\section{Collisional ionization}\label{app_cni}
We use the formulation in the code of \citet{brocklehurst1972} to obtain the values for the collisional ionization rates,
the formulation is based on \citet{burgess1968}:
\begin{eqnarray}
C_{i,n}=\frac{5.444089}{T_e^{3/2}}e^{-\chi_n} \left[ \left(\frac{5}{3}-\frac{\chi_n}{3}\right) \frac{1}{\chi_n}+\frac{1}{3}(\chi_n-1)E_1(\chi_n)e^{\chi_n}-\frac{1}{2}E_1(\chi_n)^2 e^{2\chi_n}\right],
\end{eqnarray}
in units of $\mathrm{cm^{3}~s^{-1}}$.

\end{document}